\documentclass[11pt]{article}

\usepackage[preprint]{acl}

\usepackage{times}
\usepackage{latexsym}

\usepackage[T1]{fontenc}

\usepackage[utf8]{inputenc}

\usepackage{microtype}

\usepackage{inconsolata}

\usepackage{graphicx}

\usepackage{amsmath}
\usepackage{amssymb}
\usepackage{mathtools}
\usepackage{amsthm}

\usepackage{fontawesome5}
\usepackage{hyperref}
\usepackage[utf8]{inputenc}

\usepackage{subcaption}
\usepackage{booktabs} 
\usepackage{float} 
\usepackage{multirow}

\usepackage{tikz}
\usepackage{pgfplots}
\pgfplotsset{compat=1.18}

\usepackage{colortbl} 
\definecolor{cost1st}{RGB}{255,102,102} 
\definecolor{cost2nd}{RGB}{255,153,153} 
\definecolor{cost3rd}{RGB}{255,204,204} 
\usepackage{placeins} 

\usepackage{listings}
\usepackage{titlesec}
\titlespacing*{\paragraph}{0pt}{3pt}{5pt}
\usepackage{enumitem}

\usepackage{diagbox}

\usepackage{soul}
\sethlcolor{lightgray!30}  

\definecolor{promptbg}{RGB}{248,249,250}
\definecolor{promptframe}{RGB}{206,212,218}
\definecolor{prompttext}{RGB}{33,37,41}

\usepackage{pgfplots} 
\pgfplotsset{compat=1.18}
\usepgfplotslibrary{fillbetween}

\usepackage{bbold}
\usepackage{xspace}
\newcommand{\recube}{\textsc{ReCUBE}\xspace}
\newcommand{\recubelarge}{\textsc{ReCUBE-large}\xspace}

\lstdefinestyle{promptstyle}{
  basicstyle=\ttfamily\fontsize{6}{7}\selectfont\color{prompttext},
  backgroundcolor=\color{promptbg},
  frame=single,
  rulecolor=\color{promptframe},
  breaklines=true,
  breakatwhitespace=false,
  columns=fullflexible,
  keepspaces=true,
  showstringspaces=false,
  tabsize=2,
  xleftmargin=3pt,
  xrightmargin=3pt,
  framesep=5pt,
  framexleftmargin=3pt,
  aboveskip=8pt,
  belowskip=8pt
}

\usepackage{array}
\usepackage{pifont}
\usepackage[dvipsnames]{xcolor}
\usepackage{makecell}
\usepackage{caption}
\usepackage[T1]{fontenc}
\newcommand{\cmark}{\textcolor{ForestGreen}{\ding{51}}}%
\newcommand{\xmark}{\textcolor{BrickRed}{\ding{55}}}%

\definecolor{gold1}{HTML}{E3B94C}

\definecolor{ours}{RGB}{235,245,255}

\usepackage[capitalize,noabbrev]{cleveref}

\providecommand{\nolinenumbers}{}
\providecommand{\linenumbers}{}

\title{\textsc{ReCUBE}: Evaluating Repository-Level Context Utilization in Code Generation}

\author{
  Jiseung Hong \\
  Language Technologies Institute\\
  Carnegie Mellon University \\
  \texttt{jiseungh@andrew.cmu.edu} \\\And
  Benjamin G. Ascoli \\
  Computer Science\\
  Emory University \\
  \texttt{bascoli@emory.edu} \\\And
  Jinho D. Choi \\
  Computer Science\\
  Emory University \\
  \texttt{jinho.choi@emory.edu} \\\AND
  \parbox{0.8\linewidth}{\centering\small
    \faGithub~\href{https://github.com/JiseungHong/ReCUBE}{\textbf{Code:}~\nolinkurl{github.com/JiseungHong/ReCUBE}} \\[2pt]
    \includegraphics[height=0.85em]{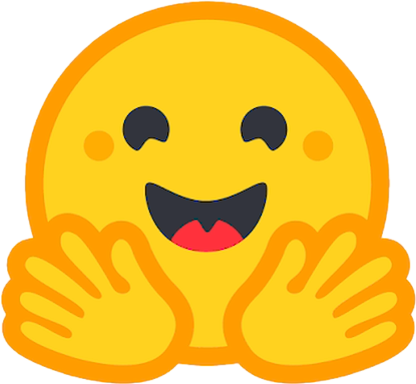}~\href{https://huggingface.co/datasets/wlqmfl1999/recube-data}{\textbf{Data:}~\nolinkurl{huggingface.co/datasets/wlqmfl1999/recube-data}} \\[2pt]
    \faDocker~\href{https://hub.docker.com/repository/docker/wlqmfl0990/recube}{\textbf{Docker:}~\nolinkurl{hub.docker.com/repository/docker/wlqmfl0990/recube}}
  }
}

\begin{document}
\maketitle
\begin{abstract}
Large Language Models (LLMs) have recently emerged as capable coding assistants that operate over large codebases through either agentic exploration or full-context generation.
Existing benchmarks capture a broad range of coding capabilities, such as resolving GitHub issues, but none of them directly isolate and measure how effectively LLMs leverage repository-level context during code generation.
To address this, we introduce \recube, a benchmark in which LLMs reconstruct a masked file within a real-world repository, using all remaining source files, dependency specifications, and documentation as their only source of context.
\recube evaluates reconstructed code with usage-aware test cases that simulate both internal module logic and external cross-file integration, reflecting real-world software usage patterns.
We further propose the Caller-Centric Exploration (CCE) toolkit, a set of dependency graph-based tools that can be integrated into agentic frameworks to guide agents toward the most relevant caller files during repository exploration.
Experiments across eight models in four settings show that repository-level context utilization remains highly challenging even for state-of-the-art models, with GPT-5 achieving only 37.57\% strict pass rate in the full-context setting.
Agents augmented with our CCE toolkit consistently outperform all baselines across all evaluated models, with improvements of up to 7.56\% in strict pass rate.
We release our benchmark, code, and evaluation framework as open source for the NLP research community.
\end{abstract}


\section{Introduction}
Over the past few years, LLM-powered code generation has shifted from passive, single-function completion to full repository-level software development~\cite{openhands, agentless, copilot, sweagent, codeplan}.
These repositories often contain thousands of lines of interdependent code spanning dozens of files, dependencies, and architectural patterns.
As a result, modern language models are required to understand, navigate, and integrate with complex, multi-file codebases.
\begin{table*}[t]
\centering
\footnotesize
\setlength{\tabcolsep}{5pt}
\renewcommand{\arraystretch}{1.2}
\begin{tabular}{l cccccc}
\toprule
\multirow{2}{*}{\textbf{Benchmark}}
  & \makecell{\textbf{Repository}\\\textbf{Level}}
  & \makecell{\textbf{Prompt-Free}\\\textbf{Evaluation}}
  & \makecell{\textbf{Contamination}\\\textbf{Free}}
  & \makecell{\textbf{Maximum}\\\textbf{Context}}
  & \makecell{\textbf{Manually}\\\textbf{Curated}}
  & \makecell{\textbf{Agent}\\\textbf{Support}} \\
\midrule
LiveCodeBench~\cite{livecodebench}
  & \xmark & \xmark & \cmark & $\sim$3K  & \cmark & \xmark \\
$\infty$Bench~\cite{infinitybench}
  & \xmark & \xmark & \xmark & 256K & \xmark & \xmark \\
SWE-Bench~\cite{swebench}
  & \cmark & \xmark & \xmark & 50K & \cmark & \cmark \\
Commit0~\cite{Commit0}
  & \cmark & \xmark & \xmark & $-$ & \cmark & \cmark \\
RepoBench-C~\cite{repobench}
  & \cmark & \cmark & \xmark & 8K & \cmark & \xmark \\
LocBench~\cite{locagent}
  & \cmark & \xmark & \xmark & $-$ & \cmark & \cmark \\
LongBench v2~\cite{longbench}\\ \quad $\hookrightarrow$ Code repo QA
  & \cmark & \xmark & \xmark & 4.1M & \cmark & \xmark \\
LongCodeBench~\cite{longcodebench}
  & \cmark & \xmark & \cmark & 1M   & \cmark & \xmark \\
LoCoBench~\cite{locobench, locobench-agent}
  & \cmark & \xmark & \cmark & 1M   & \xmark & \cmark \\
\midrule
\cellcolor{ours}\textbf{ReCUBE (Ours)} & \cellcolor{ours}\cmark & \cellcolor{ours}\cmark & \cellcolor{ours}\cmark & \cellcolor{ours}111K & \cellcolor{ours}\cmark & \cellcolor{ours}\cmark \\
\cellcolor{ours}\quad $\hookrightarrow$ \recubelarge & \cellcolor{ours}\cmark & \cellcolor{ours}\cmark & \cellcolor{ours}\cmark & \cellcolor{ours}338K & \cellcolor{ours}\cmark & \cellcolor{ours}\cmark \\
\bottomrule
\end{tabular}
\caption{\textbf{Comparison of Representative Code Benchmarks.} We compare unsaturated code generation benchmarks across six dimensions reflecting key considerations for rigorous LLM code evaluation. \textbf{Contamination Free} indicates that instances are synthetically generated, collected after known LLM training cutoffs, or otherwise explicitly filtered to exclude instances present in pretraining data.}
\label{tab:benchmark-comparison}
\vspace{-1.5em}
\end{table*}
Recent coding benchmarks implicitly require some of these capabilities, evaluating models on focused tasks such as generating libraries from scratch~\cite{Commit0} or producing test patches~\cite{swt-bench}.
Existing code localization approaches~\cite{moatless, agentless, locagent} examine how models identify relevant files and functions within a codebase, but they rely on natural language task descriptions rather than the repository structure itself.
Consequently, the community lacks a dedicated evaluation framework that isolates how effectively LLMs leverage repository-level context during code generation.

In this paper, we introduce \textbf{Re}pository-level \textbf{C}ontext \textbf{U}tilization \textbf{BE}nch (\recube), a benchmark designed to directly evaluate repository-level context utilization in code generation.
Specifically, we propose a novel task in which models are evaluated on their ability to reconstruct a masked Python file using only other source files, dependency specifications, and documentation found in the codebase as context.
As shown in Table~\ref{tab:benchmark-comparison}, \recube is a manually curated, repository-level benchmark that employs prompt-free evaluation, and scales to contexts of up to 111K tokens.
We further introduce a large-scale extension (\recubelarge) that incorporates broader codebase environments, increasing the context to 338K tokens.
To rigorously evaluate reconstructed code, we synthesize usage-aware test cases designed to simulate real-world integration by covering both internal module logic and external cross-file interactions.
We measure performance on these test cases using two metrics: Strict Pass Rate (SPR), which captures absolute correctness, and Average Pass Rate (APR), which gives credit for partial, modular functionality.

Furthermore, to support agents operating over large codebases, we propose a novel Caller-Centric Exploration (CCE) toolkit, a set of agentic tools built on a dependency graph.
It provides agents an architectural map based on incoming caller patterns, as well as interactive, graph-querying capabilities during exploration.

We evaluate eight state-of-the-art LLMs on \recube across four experimental settings, ranging from full-context generation, in which the entire task is resolved within a single-turn prompt, to interactive agentic frameworks.
The task remains highly challenging: even the best-performing model, GPT-5, achieves only a 37.57\% strict pass rate in the full-context setting.
We find that basic agentic exploration is most helpful when context windows are tight, but its advantage diminishes as context capacity increases and can eventually reverse in favor of full-context generation. 
Across all settings, agents equipped with the CCE toolkit consistently perform best, improving strict pass rate by up to $+$5.79\% over full-context generation and $+$7.56\% over vanilla agentic frameworks.



\section{Related Work}

\subsection{Repository-level Code Benchmarks}
Several benchmarks have been proposed to evaluate the repository-level coding capabilities of LLMs.
Prompt-based benchmarks assess code generation from natural language inputs, including real-world GitHub issues~\cite{swebench, swebenchverified, multiswebench} and commit messages~\cite{Commit0}.
RepoBench~\cite{repobench}, by contrast, uses cross-file dependencies as context to evaluate next-line code completion.
LongCodeBench~\cite{longcodebench}, LoCoBench~\cite{locobench, locobench-agent}, and LongBench v2~\cite{longbench} evaluate long-context code understanding through question answering over real-world GitHub issues and code repositories.

These benchmarks employ a range of evaluation metrics centered on test cases.
Many use execution-based metrics, such as resolved rate~\cite{swebench, swt-bench, longcodebench}, which measures the fraction of patches that cause all fail-to-pass tests to succeed without regressions, and unit test pass rate~\cite{Commit0}, which reports the fraction of unit tests passed by a generated implementation.
These metrics have evolved from pass@$k$~\cite{codex}, shifting toward single-attempt evaluation as repeated sampling becomes computationally prohibitive at repository scale.
In contrast, completion benchmarks rely on text-matching metrics~\cite{repobench, codexglue, intellicode} which have similarly lost favor as frontier models converge on near-perfect scores.

\subsection{Dependency Graphs for Code Generation}
Several recent software systems use dependency graphs to support a range of code generation capabilities.
\citet{rpg} construct a Repository Planning Graph and traverse it topologically to generate codebases from scratch.
\citet{codeplan} and \citet{alphatrans} use dependency graphs for repository-level code editing and translation, respectively.

Previous works primarily use graphs for retrieval, gathering dependency-aware context as an intermediate step for downstream coding tasks.
Various forms of graph-based code representations support dependency-aware retrieval for code completion via hard-coded strategies~\cite{graphcoder, grace, draco}, while more recent work exposes graph structure as an agent-queryable interface through structured queries or callable traversal tools~\cite{codexgraph, graphcodeagent, locagent}.

\begin{figure*}[t]
    \centering
    \includegraphics[width=\textwidth]{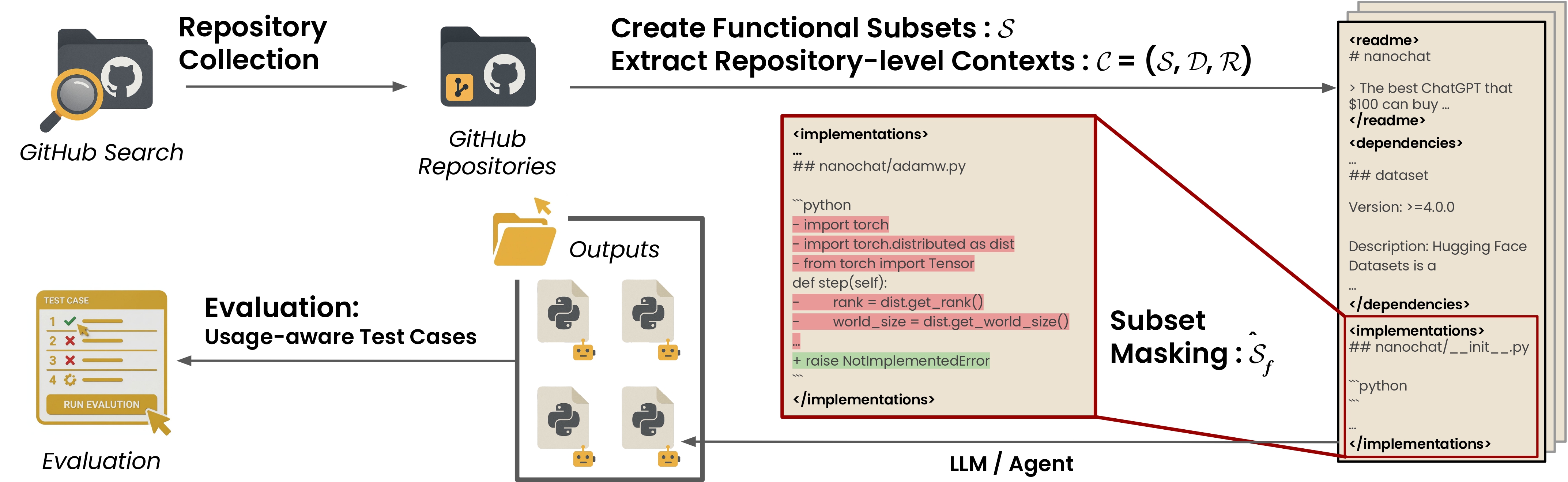} 
    \caption{\recube is constructed through the following steps: (1) \textbf{Repository Collection} from recent, high-quality GitHub repositories. (2) \textbf{Functional Subset Creation} by decomposing the repositories into file clusters that have self-contained functionality. (3) \textbf{Repository-level Context Extraction} from the functional subsets, which captures all the information needed to understand the entire subset. (4) \textbf{Subset Masking} on each functional subset, yielding a total of 366 instances.}
    \label{fig:benchmark}
    \vspace{-1.5em}
\end{figure*}

\section{\recube}
\recube is a benchmark that evaluates how well LLMs utilize repository-level context to produce functionally correct, well-integrated code.

\subsection{Benchmark Construction}
\label{ssec:bench_construction}
As illustrated in Figure~\ref{fig:benchmark}, we construct \recube through the following steps:

\paragraph{Repository Collection.} We use GitHub Search \cite{githubsearch} to collect repositories by filtering on the following criteria:
\vspace{-0.7em}
\begin{itemize}[leftmargin=3mm]
\setlength\itemsep{-0.4em}
\item Created after January 2025, the latest knowledge cutoff date among our baseline models.
\item At least 10K stars and 1K forks, using engagement as a proxy for codebase quality~\cite{zhang2025swebenchgoeslive}.
\item Released under a permissive license.
\end{itemize} 
\vspace{-0.7em}
This process yields 20 high quality repositories.

\paragraph{Functional Subset Creation.}
For each repository, we construct functional subsets ($\mathcal{S}$) by manually reviewing the codebase to identify file clusters that implement distinct, self-contained functionality.
We decompose repositories with multiple separable functionalities into up to eight subsets, and repositories containing only a singular file cluster are left in their entirety as a single subset. 
We create a total of 40 functional subsets, consisting of primarily Python source files along with highly relevant auxiliary files, such as Dockerfiles, YAML configurations, and Shell scripts. 
Detailed descriptions of each functional subset in our benchmark are provided in Appendix~\ref{appendix:functional_subsets}. 

\paragraph{Subset Masking.}
For each functional subset $\mathcal{S}$, we identify up to 10 Python files that play a central role in its implementation, resulting in a total of 366 distinct target files.
For each target file $f_i \in \mathcal{S}$, we generate a masked version of its subset, which we call an \textit{instance}, denoted as $\hat{\mathcal{S}}_{f_i}$.
This masking process involves modifying $f_i$ by removing all the import statements and replacing its function bodies with \hl{\texttt{raise NotImplementedError}}, while preserving the original signatures, docstrings, and the remaining files in $\mathcal{S}$.

\paragraph{Repository-level Context Formatting.}
For each instance $\hat{\mathcal{S}}_{f_{i}}$, we define a repository-level context $\mathcal{C}_{f_{i}}$, encompassing all available information from the repository required to reconstruct the target file $f_{i}$.
This context includes the masked instance $\hat{\mathcal{S}}_{f_{i}}$, along with the environment dependencies $D$ (including specific versions and descriptions from PyPI) and documentation files $R$ (all relevant Markdown files).
Consequently, all instances from the same functional subset $\mathcal{S}$ share much of their repository-level context but differ in their respective target file.
To assist LLMs in distinguishing these elements, we wrap each component ($\hat{S}_{f_{i}}$, $D$, $R$) using XML-style delimiter tags.
This follows established prompt engineering practices that recommend explicit structural boundaries for multi-component inputs~\cite{xml1, xml2}.

\subsection{Task Specification}
The objective of \recube is to reconstruct the masked components of a target file $f_i$ given the repository-level context $\mathcal{C}_{f_{i}}$. \recube supports evaluations for LLMs in both static and agentic settings:
\paragraph{Static (Full-Context) Setting:} 
We provide $\mathcal{C}_{f_{i}}$ as a structured prompt, including the instance $\hat{S}_{f_{i}}$, environment dependencies $D$, and documentation files $R$.
\paragraph{Agentic Setting:}
We provide fully containerized Docker and Apptainer environments.
In these settings, the instance $\hat{S}_{f_{i}}$, dependencies $D$, and documentation $R$ are installed in a manner that mirrors the original repository's file system and directory hierarchy.

\noindent Further details and illustrative examples regarding the prompts and repository-level context formatting are provided in Appendix~\ref{appendix:instance}.

\subsubsection{Test Case Generation}
\label{sec:test-case}
To rigorously evaluate the reconstructed files, we create \textit{usage-aware} unit tests for each target file using Claude Opus 4.1~\cite{anthropic2025claudecode}.
For each target file $f_i$, test cases are generated by providing a structured prompt (Appendix~\ref{appendix:test-cases}) instructing the agent to analyze both in-file and cross-file call patterns within its functional subset, generating usage-aware assertions based on real caller/callee functions.
We then validate each generated test suite within a Docker testbed, ensuring that each test passes on the gold implementation, and replace trivial or unstable tests until all cases reach sufficient difficulty and coverage across edge cases, boundary conditions, and exception handling.

\paragraph{Test Case Classification.}
Building on the in-file and cross-file patterns used during test case generation, we categorize each test case as either external or internal based on its validation focus.
Specifically, \textbf{external} test cases verify public API behavior by simulating interactions from other files in the repository, while \textbf{internal} tests focus on implementation-specific logic and private functions/methods found within the target file itself.
\recube comprises a total of 10,785 unit tests, consisting of 58.7\% external cross-file validations and 41.3\% internal in-file implementation checks (Table~\ref{tab:test_stats}).
This distribution reflects the diverse architectural roles in real-world codebases, where some modules serve as high-level APIs and others act as self-contained logic engines.

\begin{table}[htb]
\centering
\small
\begin{tabular}{lrrr}
\toprule
\textbf{Metric} & \textbf{Total} & \textbf{External} & \textbf{Internal} \\ \midrule
Test Count      & 10,785          & 6,334             & 4,451             \\
\bottomrule
\end{tabular}
\caption{
Test case statistics across 366 target files. 
}
\label{tab:test_stats}
\vspace{-0.5em}
\end{table}


\subsubsection{Evaluation Metrics}
We report two primary metrics to evaluate model performance. Let $T_i$ represent the total number of test cases and $P_i$ the number of tests passing for a given instance $i$, across a total of $N$ instances. We define Strict Pass Rate (SPR) and Average Pass Rate (APR) as follows:

\vspace{-2em}

\begin{align}
&\text{SPR} = \frac{1}{N} \sum_{i=1}^{N} \mathbb{1}[P_i = T_i] \times 100 \\
&\text{APR} = \frac{1}{N} \sum_{i=1}^{N} \frac{P_i}{T_i} \times 100
\end{align}

\vspace{-2pt}

\noindent where $\mathbb{1}(\cdot)$ denotes the indicator function, which evaluates to 1 if the condition is met and 0 otherwise. These metrics are analogous to \textit{Strict Accuracy} and \textit{Test Case Average}, respectively, as proposed in the APPS benchmark~\cite{apps}.

Prior benchmarks primarily evaluate correctness based only on binary outcome per issue (SPR).
We additionally report Average Pass Rate because in \recube, each test case independently exercises a distinct usage context.
Partial credit thus reflects how well the models implement modularized functionalities.

\subsection{Features of \recube}

\subsubsection{Functional Domains}
\label{sec:domains}

\begin{figure}[H]
\centering
\begin{tikzpicture}[scale=0.85]
\definecolor{agent}{HTML}{4E79A7}
\definecolor{tool}{HTML}{F28E2B}
\definecolor{llm}{HTML}{E15759}
\definecolor{voice}{HTML}{76B7B2}
\definecolor{rag}{HTML}{59A14F}
\definecolor{code}{HTML}{EDC948}

\def\radius{2.2}
\def\innerradius{1.1}

\fill[agent] (0,0) -- (90:\radius) arc (90:18:\radius) -- cycle;
\fill[tool] (0,0) -- (18:\radius) arc (18:-81:\radius) -- cycle;
\fill[llm] (0,0) -- (-81:\radius) arc (-81:-162:\radius) -- cycle;
\fill[voice] (0,0) -- (-162:\radius) arc (-162:-198:\radius) -- cycle;
\fill[rag] (0,0) -- (-198:\radius) arc (-198:-225:\radius) -- cycle;
\fill[code] (0,0) -- (-225:\radius) arc (-225:-270:\radius) -- cycle;

\fill[white] (0,0) circle (\innerradius);

\node[font=\small\bfseries] at (0,0.15) {40};
\node[font=\tiny] at (0,-0.15) {variations};

\def\labelradius{2.9}

\draw[gray!60, thin] (54:\radius) -- (54:\labelradius);
\node[font=\scriptsize, anchor=west] at (54:\labelradius+0.05) {Agents (8)};

\draw[gray!60, thin] (-31:\radius) -- (-31:\labelradius);
\node[font=\scriptsize, anchor=west] at (-31:\labelradius+0.05) {Tool/MCP (11)};

\draw[gray!60, thin] (-121:\radius) -- (-121:\labelradius);
\node[font=\scriptsize, anchor=east] at (-121:\labelradius+0.05) {LLM (9)};

\draw[gray!60, thin] (-180:\radius) -- (-180:\labelradius);
\node[font=\scriptsize, anchor=east] at (-180:\labelradius+0.05) {Voice (4)};

\draw[gray!60, thin] (-211:\radius) -- (-211:\labelradius);
\node[font=\scriptsize, anchor=east] at (-211:\labelradius+0.05) {RAG (3)};

\draw[gray!60, thin] (-247:\radius) -- (-247:\labelradius);
\node[font=\scriptsize, anchor=east] at (-247:\labelradius+0.05) {Code/Vision (5)};

\end{tikzpicture}
\caption{Distribution of benchmark variations across functional domains.}
\label{fig:domain_distribution}
\end{figure}
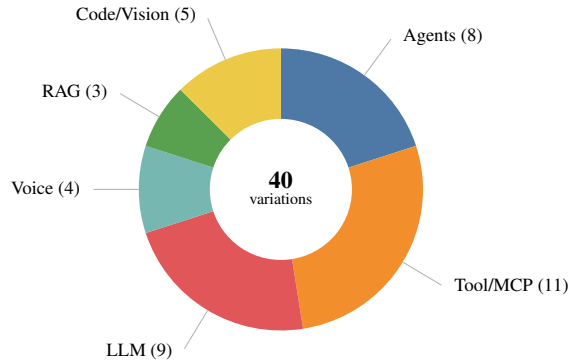

We categorize the 40 functional subsets (\textsection\ref{ssec:bench_construction}) into six functional domains: Tool \& MCP Integration, LLM Services, Agent Frameworks, Code \& Vision, Voice \& Speech, and RAG (Figure~\ref{fig:domain_distribution}).
This distribution reflects the current AI engineering landscape, where LLM related tool integration dominates development efforts.

\subsubsection{Docstring Completeness}
\begin{figure}[H]
\centering
\begin{tikzpicture}
\begin{axis}[
    ybar,
    width=0.95\columnwidth,
    height=5cm,
    bar width=14pt,
    ylabel={Instances (\%)},
    symbolic x coords={Detailed, Args/Ret., Signature, None},
    xtick=data,
    x tick label style={font=\footnotesize},
    y tick label style={font=\footnotesize},
    ylabel style={font=\footnotesize},
    ymin=0, ymax=60,
    ytick={0,10,20,30,40,50,60},
    enlarge x limits=0.18,
    axis lines*=left,
    ymajorgrids=true,
    grid style={dashed, gray!30},
    legend style={
        at={(0.5,-0.22)},
        anchor=north,
        font=\footnotesize,
        draw=none,
        fill=none,
        legend columns=2,
        column sep=8pt,
    },
    legend cell align={left},
    nodes near coords={%
        \pgfmathfloatifflags{\pgfplotspointmeta}{0}{}{\pgfmathprintnumber\pgfplotspointmeta\%}%
    },
    nodes near coords style={font=\scriptsize, anchor=south},
]
\addplot[fill=blue!65, draw=blue!75, bar shift=0pt] coordinates
    {(Detailed,10.9) (Args/Ret.,7.2) (Signature,27.5) (None,0)};
\addplot[fill=gray!50, draw=gray!60, bar shift=0pt] coordinates
    {(Detailed,0) (Args/Ret.,0) (Signature,0) (None,54.4)};
\legend{Documented (45.6\%), Undocumented (54.4\%)}
\end{axis}
\end{tikzpicture}
\vspace{-0.5em}
\caption{Distribution of documentation completeness.}
\label{fig:docstring}
\end{figure}
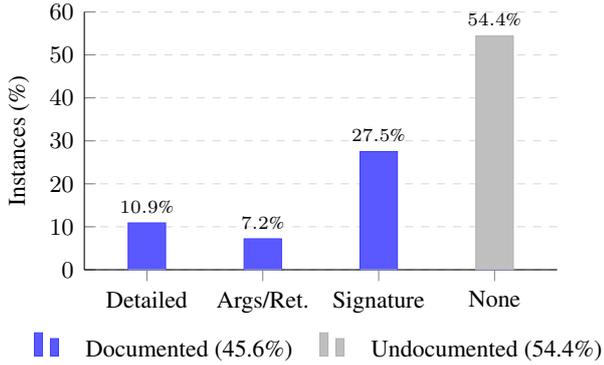

Figure~\ref{fig:docstring} shows the distribution of docstring completeness in target files across four categories. 
The majority (54.5\%) of the target files lack documentation entirely, while 45.6\% have varying degrees of documentation completeness.
27.5\% contain only function signatures with a one-line description, 7.2\% include argument and return value descriptions, and the remaining 10.9\% feature detailed docstrings, with arguments, return values, and examples.
This distribution reflects real-world codebases, where documentation is usually incomplete, requiring models to infer functionality from other parts of the repository.
Further details and examples for each docstring category can be found in Appendix~\ref{appendix:docstrings}.

\vspace{-1em}

\subsection{Large-Scale Extension}
\label{sec:large-scale}

To supplement \recube with larger repository-level context scenarios and investigate LLM and agent behavior under longer context conditions, we introduce \recubelarge, an extended evaluation set.
We construct \recubelarge by merging functional subsets that originate from the same source repository, yielding 6 large-scale repositories that provide complete codebase context spanning multiple functional modules.
\recubelarge uses the same target files from the original functional subsets, resulting in 138 instances that maintain the same task but provide significantly more context.
Compared to \recube, \recubelarge has a significantly increased average input context, and a maximum context of over 3$\times$ (Table~\ref{tab:large-spec}).

\begin{table}[htb]
\centering
\small
\begin{tabular}{lccc}
\toprule
\textbf{Dataset} & \textbf{\makecell{Average\\Context}} & \textbf{\makecell{Maximum\\Context}} & \textbf{\# Instances} \\
\midrule
\recube      & 71K  & 111K & 366 \\
\recubelarge & 146K & 338K & 138 \\
\bottomrule
\end{tabular}
\caption{Statistics for \recube and \recubelarge.}
\label{tab:large-spec}
\vspace{-1em}
\end{table}

\section{Experiments}

We conduct a systematic evaluation on \recube across four experimental settings, including three established baselines and a fourth proposed method, all using a diverse set of LLMs.
To prevent data contamination, we disable built-in tool capabilities, such as \hl{\texttt{web\_search}}, in all configurations.
Comprehensive configuration details for LLMs and agentic frameworks are provided in Appendix~\ref{appendix:config}.

\subsection{Baselines}
\label{ssec:baselines}
\paragraph{Full-Context.} For each target file $f_i$, models receive the full repository-level context $\mathcal{C}_{f_{i}}$ in a single prompt and reconstruct the target file directly. This setting serves as our primary baseline, testing code generation capabilities when all necessary information is provided upfront.
We evaluate eight models spanning different capability tiers: Claude 3 Haiku~\cite{claude3}, Gemini 2.5 Flash and Gemini 2.5 Pro~\cite{gemini25}, GPT-5 and GPT-5 Mini~\cite{gpt5}, GPT-OSS 20B~\cite{gptoss}, Devstral Small 2507~\cite{devstral}, and Qwen3-Coder 30B~\cite{qwen3}.

\paragraph{Full-Context + CoT.} To assess whether explicit reasoning improves generation quality, we prompt models to reason step-by-step~\cite{cot, cot-codegen} before generating the target file. This encourages planning of imports, dependencies, and implementation logic prior to code generation. We evaluate GPT-5, GPT-5 Mini, GPT-OSS 20B, Devstral Small 2507, and Qwen3-Coder 30B in this setting.

\paragraph{Agent.} We employ an agentic framework based on Mini SWE-agent~\cite{sweagent}, often called a bash-only agent, that enables models to operate within an interactive environment, allowing them to explore the repository using standard bash commands while reconstructing the target file.
In contrast to the Full-Context setting, the agent relies entirely on the LLM’s capability to identify relevant information from the repository.
We evaluate this approach across GPT-5 Mini, Devstral Small 2507, and Qwen3-Coder 30B.

\subsection{Eliciting Better Context Utilization in Agents}
\label{ssec:graph}

In addition to the baselines, we propose a novel toolkit that uses a dependency graph to guide agents to the most relevant context during code reconstruction.

\subsubsection{Exploring Callers Benefits Hard Tasks}
\label{sssec:motivation_method}
To understand whether agent performance is driven by caller coverage (the percentage of files invoking target functions that the agent views), we investigate Kendall-Tau~\cite{kendal-tau} and Spearman~\cite{spearman} correlations between caller coverage and APR, following previous research~\cite{llm-kendal-tau, code-pearson1, code-pearson2}. 
We analyze GPT-5 Mini and Qwen3-Coder 30B in the bash-only agent baseline setting (\S\ref{ssec:baselines}), stratifying instances into per-model difficulty quartiles based on their performance.
To prevent statistical skew from outliers, we exclude both trivial cases (100\% pass rate) and unsolvable cases (0\% pass rate) from this configuration.

\begin{table}[htb]
\centering
\small
\setlength{\tabcolsep}{3pt}
\begin{tabular}{@{}l@{\hspace{18pt}}r@{\hspace{4pt}}c@{\hspace{18pt}}r@{\hspace{4pt}}c@{}}
\toprule
& \multicolumn{2}{c}{\textbf{GPT-5 Mini}}
& \multicolumn{2}{c}{\textbf{Qwen3-Coder 30B}} \\
\cmidrule(lr){2-3}\cmidrule(lr){4-5}
\textbf{Difficulty}
  & \multicolumn{1}{c}{$\tau$} & \multicolumn{1}{c}{$\rho$}
  & \multicolumn{1}{c}{$\tau$} & \multicolumn{1}{c}{$\rho$} \\
\midrule
Hardest & $-$0.184 & $-$0.259 & $+$0.020 & $+$0.026 \\
Hard    & $+$0.280* & $+$0.404* & $+$0.304* & $+$0.408* \\
Medium  & $+$0.125 & $+$0.172 & $-$0.171 & $-$0.217 \\
Easiest & $-$0.063 & $-$0.086 & $-$0.135 & $-$0.200 \\
\midrule
Overall & $+$0.009 & $+$0.013 & $-$0.073 & $-$0.106 \\
\bottomrule
\multicolumn{5}{@{}l}{\scriptsize $\tau$: Kendall-Tau; $\rho$: Spearman. *$p < 0.05$; all others $p > 0.05$.}
\end{tabular}
\vspace{-5pt}
\caption{Correlations between caller coverage and Average Pass Rate by task difficulty.}
\label{tab:coverage_correlation}
\vspace{-1em}
\end{table}

The results of this analysis are presented in Table~\ref{tab:coverage_correlation}.
We find a significant positive correlation between caller coverage and pass rate on hard tasks ($\rho=+0.408, p=0.035$), suggesting that agents benefit from viewing files that invoke target functions.
These findings show that navigating agents toward most relevant caller files significantly aids target file reconstruction.

\subsubsection{Caller-Centric Exploration Toolkit}
\label{sssec:method}

Towards this end, we introduce the \textit{Caller-Centric Exploration (CCE)} toolkit, a suite of agentic tools designed to enhance repository-level navigation by integrating dependency graph analysis directly into the agent’s exploration loop.
The CCE toolkit helps models establish an exploration roadmap by surfacing incoming caller patterns, specifically identifying which files invoke or import the target code, along with call hierarchies and contextual relationships prior to implementation.
Beyond this capability, the toolset also provides actionable features, enabling the agent to interactively query the repository as a heterogeneous graph for granular context.
This dependency-aware perspective allows agents to explore the codebase while ensuring that choices are informed by a clear understanding of the target file's usage and integration points.

\paragraph{Dependency Graphs for Codebase.}
We index each instance $\hat{\mathcal{S}}_{f_{i}}$ as a single heterogeneous directed graph following \citet{locagent}, capturing four entity types (directory, file, class, function) and four relationship types (contain, import, invoke, inherit).
Notably, the target file $f_{i}$ is withheld from each of the graph to prevent access to masked information.

\paragraph{Tool specifications.}
The Caller-Centric Exploration toolkit contains three tools that enable agents to acquire a deeper understanding of the role a target file $f_i$ performs within the instance $\hat{\mathcal{S}}_{f_{i}}$.
The primary tool provides the agent with extensive information across four categories prior to exploration:

\vspace{-0.7em}
\begin{itemize}[leftmargin=3mm]
\setlength\itemsep{-0.4em}
    \item \textbf{Caller Patterns} identify which functions, methods, and classes invoke the target code and their respective call frequencies.
    \item \textbf{Inheritance} shows class hierarchies to illustrate how the target file's classes relate to their parents and siblings, clarifying code architecture.
    \item \textbf{Module Context} reveals directory organization by surfacing neighboring files and their shared imports.
    \item \textbf{Similar Files} identifies files with comparable structures by aggregating 3 distinct similarity signals: \textit{Filename} uses subtoken overlap on filenames, calculated on underscore-separated segments; \textit{Structure} compares AST-extracted counts of classes, functions, and nesting depth to quantify hierarchical alignment; \textit{Identifier} computes the normalized BM25 score of extracted code identifiers (e.g., function, class names) across the repository corpus.
\end{itemize}
\vspace{-0.7em}

\noindent To further support the agent during codebase exploration, we provide a second tool for programmatically querying the dependency graph via inline Python scripts to perform custom structural analysis.
For example, agents can execute a multi-line scripts to retrieve specific entity interactions and relations.
Finally, we provide a validation tool that allows agents to perform syntax checks, ensuring that all modifications are syntactically correct before finalization. 
Further specifications and examples can be found in Appendix~\ref{appendix:caller-first-principle}.

\subsection{Agent + CCE toolkit}
\label{ssec:agent_cce}
We extend the base agentic framework by implementing the CCE toolkit (\S\ref{sssec:method}), which augments the agent with dependency graph-based tools to strategically guide repository exploration.
By surfacing relevant caller files and exposing the underlying repository structure, we aim to mitigate the limited contextual navigation observed in basic agentic approach.
We use the same models as in the baseline agent setting (\S\ref{ssec:baselines}).

\section{Analysis}
\begin{table*}[t]
\centering
\small
\begin{tabular}{ll cccc cc}
\toprule
& & \multicolumn{2}{c}{\textbf{Overall (\%)}} & \multicolumn{2}{c}{\textbf{External TCs (\%)}} & \multicolumn{2}{c}{\textbf{Internal TCs (\%)}} \\
\cmidrule(lr){3-4} \cmidrule(lr){5-6} \cmidrule(lr){7-8}
\textbf{Setting} & \textbf{Model} & \textbf{SPR} & \textbf{APR} & \textbf{SPR} & \textbf{APR} & \textbf{SPR} & \textbf{APR} \\
\midrule
\midrule
\multirow{8}{*}{Full-Context}
 & GPT-OSS 20B           & 25.07$\:\:$$\:\:$$\:\:$$\:\:$ & 36.83$\:\:$$\:\:$$\:\:$$\:\:$ & 22.19$\:\:$$\:\:$$\:\:$$\:\:$ & 32.61$\:\:$$\:\:$$\:\:$$\:\:$ & 30.63$\:\:$$\:\:$$\:\:$$\:\:$ & 39.11$\:\:$$\:\:$$\:\:$$\:\:$ \\
 & Devstral Small 2507    & 21.76$\:\:$$\:\:$$\:\:$$\:\:$ & 30.20$\:\:$$\:\:$$\:\:$$\:\:$ & 17.36$\:\:$$\:\:$$\:\:$$\:\:$ & 25.57$\:\:$$\:\:$$\:\:$$\:\:$ & 27.93$\:\:$$\:\:$$\:\:$$\:\:$ & 34.99$\:\:$$\:\:$$\:\:$$\:\:$ \\
 & Qwen3-Coder 30B       & 25.34$\:\:$$\:\:$$\:\:$$\:\:$ & 38.54$\:\:$$\:\:$$\:\:$$\:\:$ & 21.86$\:\:$$\:\:$$\:\:$$\:\:$ & 34.44$\:\:$$\:\:$$\:\:$$\:\:$ & 31.23$\:\:$$\:\:$$\:\:$$\:\:$ & 40.88$\:\:$$\:\:$$\:\:$$\:\:$ \\
 \cmidrule(lr){2-8}
 & Claude 3 Haiku         & 18.66$\:\:$$\:\:$$\:\:$$\:\:$ & 23.17$\:\:$$\:\:$$\:\:$$\:\:$ & 14.79$\:\:$$\:\:$$\:\:$$\:\:$ & 18.85$\:\:$$\:\:$$\:\:$$\:\:$ & 21.02$\:\:$$\:\:$$\:\:$$\:\:$ & 25.74$\:\:$$\:\:$$\:\:$$\:\:$ \\
 & Gemini 2.5 Flash       & 25.69$\:\:$$\:\:$$\:\:$$\:\:$ & 40.28$\:\:$$\:\:$$\:\:$$\:\:$ & 23.15$\:\:$$\:\:$$\:\:$$\:\:$ & 37.26$\:\:$$\:\:$$\:\:$$\:\:$ & 33.03$\:\:$$\:\:$$\:\:$$\:\:$ & 43.00$\:\:$$\:\:$$\:\:$$\:\:$ \\
 & Gemini 2.5 Pro         & 29.56$\:\:$$\:\:$$\:\:$$\:\:$ & 51.53$\:\:$$\:\:$$\:\:$$\:\:$ & 27.97$\:\:$$\:\:$$\:\:$$\:\:$ & 48.61$\:\:$$\:\:$$\:\:$$\:\:$ & 40.24$\:\:$$\:\:$$\:\:$$\:\:$ & 54.14$\:\:$$\:\:$$\:\:$$\:\:$ \\
 & GPT-5 Mini            & 33.43$\:\:$$\:\:$$\:\:$$\:\:$ & 54.02$\:\:$$\:\:$$\:\:$$\:\:$ & 31.51$\:\:$$\:\:$$\:\:$$\:\:$ & 50.23$\:\:$$\:\:$$\:\:$$\:\:$ & 40.84$\:\:$$\:\:$$\:\:$$\:\:$ & 56.20$\:\:$$\:\:$$\:\:$$\:\:$ \\
 & GPT-5                & 37.57$\:\:$$\:\:$$\:\:$$\:\:$& \textbf{60.43}$\:\:$$\:\:$$\:\:$$\:\:$ & 35.37$\:\:$$\:\:$$\:\:$$\:\:$ & \textbf{57.24}$\:\:$$\:\:$$\:\:$$\:\:$ & 45.65$\:\:$$\:\:$$\:\:$$\:\:$ & \textbf{63.74}$\:\:$$\:\:$$\:\:$$\:\:$ \\
\midrule
\multirow{5}{*}{\begin{tabular}[l]{@{}l@{}}Full-Context\\ {\scriptsize + CoT}\end{tabular}}
 & GPT-OSS 20B           & 26.59\textcolor{blue}{\textsubscript{+1.52}} & 38.67\textcolor{blue}{\textsubscript{+1.84}} & 23.15\textcolor{blue}{\textsubscript{+0.96}} & 34.93\textcolor{blue}{\textsubscript{+2.32}} & 33.03\textcolor{blue}{\textsubscript{+2.40}} & 41.61\textcolor{blue}{\textsubscript{+2.50}} \\
 & Devstral Small 2507    & 19.22\textcolor{red}{\textsubscript{-2.54}} & 25.28\textcolor{red}{\textsubscript{-4.92}} & 15.11\textcolor{red}{\textsubscript{-2.25}} & 19.65\textcolor{red}{\textsubscript{-5.92}} & 21.32\textcolor{red}{\textsubscript{-6.61}} & 27.68\textcolor{red}{\textsubscript{-7.31}} \\
 & Qwen3-Coder 30B       & 24.86\textcolor{red}{\textsubscript{-0.48}} & 38.05\textcolor{red}{\textsubscript{-0.49}} & 22.51\textcolor{blue}{\textsubscript{+0.65}} & 34.09\textcolor{red}{\textsubscript{-0.35}} & 31.53\textcolor{blue}{\textsubscript{+0.30}} & 41.58\textcolor{blue}{\textsubscript{+0.70}} \\
 \cmidrule(lr){2-8}
 & GPT-5 Mini            & 33.98\textcolor{blue}{\textsubscript{+0.55}} & 55.50\textcolor{blue}{\textsubscript{+1.48}} & 32.15\textcolor{blue}{\textsubscript{+0.64}} & 51.63\textcolor{blue}{\textsubscript{+1.40}} & 43.54\textcolor{blue}{\textsubscript{+2.70}} & 58.62\textcolor{blue}{\textsubscript{+2.42}} \\
 & GPT-5                & 37.29\textcolor{red}{\textsubscript{-0.28}} & 58.98\textcolor{red}{\textsubscript{-1.45}} & 34.41\textcolor{red}{\textsubscript{-0.96}} & 54.98\textcolor{red}{\textsubscript{-2.26}} & \textbf{45.95}\textcolor{blue}{\textsubscript{+0.30}} & 61.29\textcolor{red}{\textsubscript{-2.45}} \\

\midrule
\multirow{3}{*}{\begin{tabular}[l]{@{}l@{}}Agent\end{tabular}}
 & Devstral Small 2507    & 23.82\textcolor{blue}{\textsubscript{+2.06}} & 35.21\textcolor{blue}{\textsubscript{+5.01}} & 20.26\textcolor{blue}{\textsubscript{+2.90}} & 29.35\textcolor{blue}{\textsubscript{+3.78}} & 30.33\textcolor{blue}{\textsubscript{+2.40}} & 40.83\textcolor{blue}{\textsubscript{+5.84}} \\
 & Qwen3-Coder 30B       & 25.76\textcolor{blue}{\textsubscript{+0.42}} & 38.94\textcolor{blue}{\textsubscript{+0.40}} & 20.26\textcolor{red}{\textsubscript{-1.60}} & 21.86\textcolor{red}{\textsubscript{-12.58}} & 33.33\textcolor{blue}{\textsubscript{+2.10}} & 44.45\textcolor{blue}{\textsubscript{+3.57}} \\
 \cmidrule(lr){2-8}
 & GPT-5 Mini            & 31.22\textcolor{red}{\textsubscript{-2.21}} & 50.56\textcolor{red}{\textsubscript{-3.46}} & 28.62\textcolor{red}{\textsubscript{-2.89}} & 47.31\textcolor{red}{\textsubscript{-2.92}} & 38.44\textcolor{red}{\textsubscript{-2.40}} & 54.00\textcolor{red}{\textsubscript{-2.20}} \\
\midrule
\multirow{3}{*}{\begin{tabular}[l]{@{}l@{}}Agent\\ {\scriptsize + CCE toolkit (ours)}\end{tabular}}
 & Devstral Small 2507    & 25.76\textcolor{blue}{\textsubscript{+4.00}} & 38.45\textcolor{blue}{\textsubscript{+8.25}} & 22.19\textcolor{blue}{\textsubscript{+4.83}} & 33.62\textcolor{blue}{\textsubscript{+8.05}} & 31.83\textcolor{blue}{\textsubscript{+3.90}} & 43.23\textcolor{blue}{\textsubscript{+8.24}} \\
 & Qwen3-Coder 30B       & 28.45\textcolor{blue}{\textsubscript{+3.11}} & 43.36\textcolor{blue}{\textsubscript{+4.82}} & 25.08\textcolor{blue}{\textsubscript{+3.22}} & 38.77\textcolor{blue}{\textsubscript{+4.33}} & 36.34\textcolor{blue}{\textsubscript{+5.11}} & 48.11\textcolor{blue}{\textsubscript{+7.23}} \\
 \cmidrule(lr){2-8}
 & GPT-5 Mini            & \textbf{38.78}\textcolor{blue}{\textsubscript{+5.35}} & 58.39\textcolor{blue}{\textsubscript{+4.37}} & \textbf{37.30}\textcolor{blue}{\textsubscript{+5.79}} & 55.16\textcolor{blue}{\textsubscript{+4.93}} & 45.35\textcolor{blue}{\textsubscript{+4.51}} & 60.16\textcolor{blue}{\textsubscript{+3.96}} \\
\bottomrule
\end{tabular}
\caption{\textbf{Experimental Results on \textsc{ReCUBE}.} Results show overall Strict Pass Rate (SPR) and Average Pass Rate (APR) along with External and Internal Test Cases (TCs) pass rates. Subscripts denote change from the full-context baseline.}
\label{tab:main_results}
\vspace{-1.75em}
\end{table*}

The full results of our experiments are shown in Table~\ref{tab:main_results}.
GPT-5 performs best in the full-context setting (37.57\% SPR, 60.43\% APR), but smaller open-source models demonstrate proficiency comparable to the weaker closed-source models, with Qwen3-Coder achieving a 25.34\% SPR.
Internal test cases consistently achieve higher performance across all models ($+$9.55\% SPR on average compared to external), suggesting that reconstructing self-contained functionality is easier than implementing cross-file integration.

Chain-of-Thought (CoT) reasoning degrades the performance of GPT-5, Devstral, and Qwen3-Coder.
Since GPT-5 already generates internal reasoning traces, explicit CoT may be redundant.
For Devstral and Qwen3-Coder, CoT consumes valuable context window space without yielding observable benefits, consistent with prior findings~\cite{codecot}.
However, CoT does provide moderate gains for GPT-OSS and GPT-5 Mini, with the former likely due to its design that explicitly supports structured chain-of-thought generation~\cite{gptoss, codescot}.

\noindent While basic agentic exploration yields gains in overall SPR and APR for models with limited context capacity, these improvements diminish as capacity grows.
Devstral Small, with a context window of 131K tokens, gains 2.06\%/5.01\% SPR/APR, while Qwen3-Coder, which has a higher context window of 256K tokens, gains only 0.42\%/0.40\% SPR/APR.
GPT-5 Mini, with a 400K token context limit, actually performs better in the full-context setting, losing about 2.21\%/3.46\% SPR/APR with the agent setting.

However, when equipped with our proposed CCE toolkit, agents dramatically improve, consistently outperforming every other framework across all models and metrics, improving on full-context settings by up to 8.24\% for Devstral Small on internal APR and 5.79\% for GPT-5 Mini on external SPR. 
With the CCE toolkit, GPT-5 Mini is able achieve the best performance on overall SPR, even outperforming GPT-5 using full-context, a dramatic 5.35\% improvement.

\paragraph{Domain-based Performance Analysis.}

\begin{figure*}[t]
  \centering
  \begin{tikzpicture}
    \begin{axis}[
        ybar,
        bar width        = 0.13cm,
        width            = \textwidth,
        height           = 5.5cm,
        enlarge x limits = 0.08,
        ylabel           = {Strict Pass Rate (\%)},
        ylabel style     = {font=\footnotesize, yshift=2pt},
        ymin = 0, ymax   = 55,
        ytick            = {0,10,20,30,40,50},
        yticklabel style = {font=\scriptsize},
        xtick            = data,
        xticklabels      = {
            Agent\\Frameworks,
            LLM\\Services,
            Tool \& MCP\\Integration,
            RAG,
            Voice \&\\Speech,
            Code \&\\Vision
        },
        xticklabel style = {
            align      = center,
            font       = \scriptsize,
            text width = 2.0cm,
        },
        legend style = {
            at             = {(0.5, -0.23)},
            anchor         = north,
            legend columns = 4,
            font           = \scriptsize,
            draw           = none,
            fill           = none,
            column sep     = 10pt,
            row sep        = 2pt,
        },
        legend cell align = {left},
        legend image code/.code = {
            \draw[#1, draw opacity=1]
              (0.00cm, 0.00cm) rectangle (0.14cm, 0.28cm);
            \draw[#1, draw opacity=1]
              (0.18cm, 0.00cm) rectangle (0.32cm, 0.18cm);
        },
        ymajorgrids  = true,
        grid style   = {dashed, gray!25},
        axis lines*  = left,
        tick align   = outside,
        tick style   = {color=black!60},
        axis line style = {color=black!60},
    ]

    \addplot[fill=cyan!45!white, draw=cyan!65!black, line width=0.4pt]
        coordinates {
            (1, 36.23) (2, 44.19) (3, 41.11)
            (4, 46.67) (5, 22.50) (6, 27.66)
        };
    \addlegendentry{GPT-5}

    \addplot[fill=blue!60!white, draw=blue!80!black, line width=0.4pt]
        coordinates {
            (1, 34.78) (2, 44.19) (3, 35.56)
            (4, 33.33) (5, 17.50) (6, 21.28)
        };
    \addlegendentry{GPT-5 mini}

    \addplot[fill=black!75!white, draw=black, line width=0.4pt]
        coordinates {
            (1, 36.23) (2, 38.37) (3, 30.00)
            (4, 26.67) (5, 10.00) (6, 21.28)
        };
    \addlegendentry{Gemini 2.5 Pro}

    \addplot[fill=gray!45!white, draw=gray!70!black, line width=0.4pt]
        coordinates {
            (1, 27.54) (2, 33.72) (3, 23.33)
            (4, 31.03) (5, 12.50) (6, 20.83)
        };
    \addlegendentry{Gemini 2.5 Flash}

    \addplot[fill=orange!65!white, draw=orange!85!black, line width=0.4pt]
        coordinates {
            (1, 30.43) (2, 36.05) (3, 23.33)
            (4, 20.00) (5,  7.50) (6, 20.83)
        };
    \addlegendentry{Qwen3-Coder 30B}

    \addplot[fill=purple!55!white, draw=purple!75!black, line width=0.4pt]
        coordinates {
            (1, 30.43) (2, 37.21) (3, 27.78)
            (4, 20.69) (5, 10.00) (6, 17.02)
        };
    \addlegendentry{GPT-OSS-20B}

    \addplot[fill=red!55!white, draw=red!75!black, line width=0.4pt]
        coordinates {
            (1, 21.74) (2, 33.72) (3, 22.22)
            (4, 16.67) (5,  0.50) (6, 20.83)
        };
    \addlegendentry{Devstral Small 2507}

    \addplot[fill=green!55!white, draw=green!70!black, line width=0.4pt]
        coordinates {
            (1, 21.74) (2, 31.40) (3, 15.56)
            (4, 10.71) (5,  0.50) (6, 17.02)
        };
    \addlegendentry{Claude 3 Haiku}

    \end{axis}
  \end{tikzpicture}
  \vspace{-0.5em}
  \caption{%
    \textbf{Performance Across Functional Domains.} We report Strict Pass Rate (\%) across six functional domains for eight models in the full-context setting.
  }
  \vspace{-1em}
  \label{fig:domain_performance}
\end{figure*}
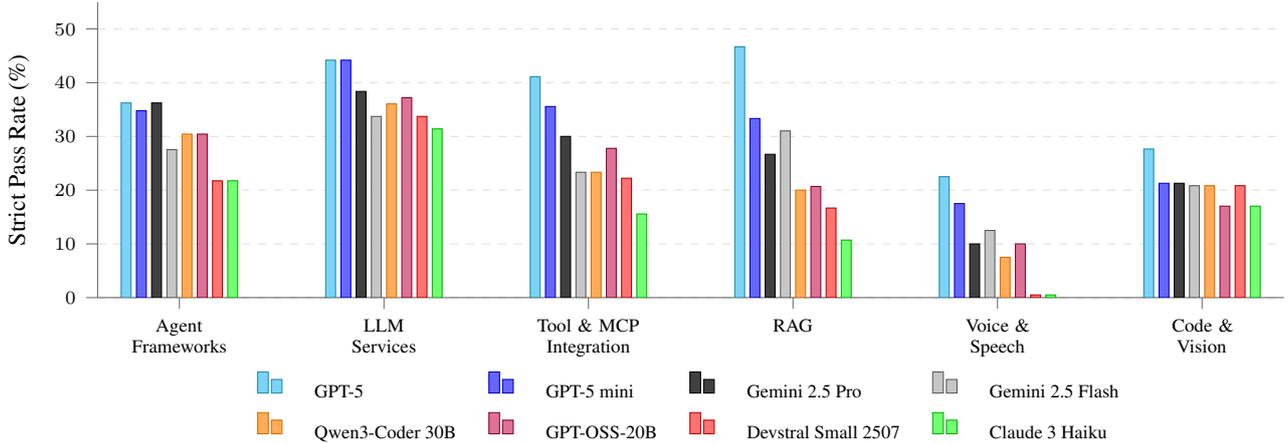

We analyze performance across the six functional domains (\S\ref{sec:domains}) in the Full-Context setting in Figure~\ref{fig:domain_performance}.
In general, LLM-centric domains such as LLM services and agent frameworks demonstrate generally higher tractability, with best-performing models achieving 44.19\% and
36.23\% SPR respectively.
In contrast, tasks requiring multimodal understanding, particularly voice and speech processing, impose substantially greater difficulty, and models struggle to achieve strong results.

\paragraph{Caller Coverage Analysis on Agent Trajectories.}

Using GPT-5 Mini and Qwen3-Coder, we analyze the caller coverage of our CCE-augmented agent compared to the bash-only baseline, and whether this translates into performance gains, stratified by task difficulty.
We use the same per-model difficulty quartiles introduced in \S~\ref{sssec:motivation_method}, but this time include the outliers with 0\% and 100\% pass rate.
CCE consistently improves both models' caller coverage across all quartiles, with GPT-5 Mini showing the largest gain (Figure~\ref{fig:absolute_quartiles}).
On SPR, however, the gains are concentrated on harder tasks, and the biggest gain occurs on instances that had 0\% SPR (the hardest tasks) in the bash-only agentic setting, with both models peaking at the hard and hardest quartiles.
This pattern is consistent with the initial correlation analysis: on easy tasks, agents can already solve instances with limited caller context, so additional coverage yields marginal (or slightly negative) SPR returns despite consistent coverage gains.
Conversely, on harder tasks, where caller relationships are critical for correct reconstruction, the improved coverage from our CCE toolkit translates into significant performance improvements. 

\usetikzlibrary{patterns}

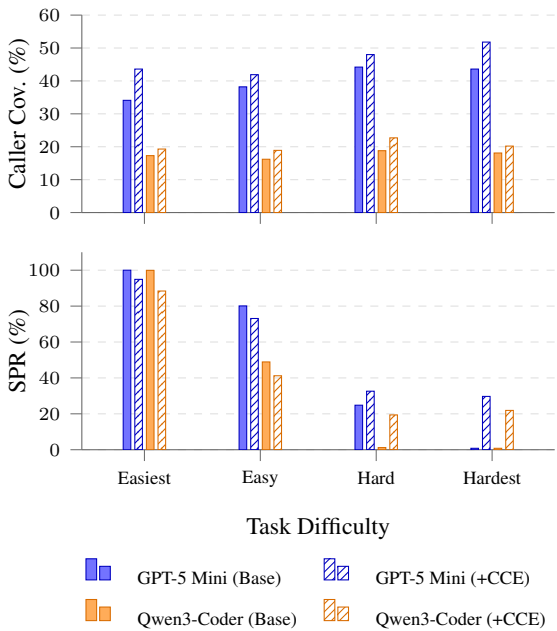
\begin{figure}[t!]
\centering

{\hfuzz=50pt
\begin{tikzpicture}[trim axis left, trim axis right]
\begin{axis}[
  ybar, bar width=0.10cm, ybar=1.5pt,
  width=0.95\columnwidth,
  height=4.2cm,
  enlarge x limits=0.18,
  ylabel={Caller Cov. (\%)},
  ylabel style={font=\footnotesize, yshift=-4pt},
  ymin=0, ymax=60,
  ytick={0,10,20,30,40,50,60},
  yticklabel style={font=\scriptsize},
  xtick=data,
  symbolic x coords={Q1,Q2,Q3,Q4},
  xticklabels={,,,,},
  ymajorgrids=true,
  grid style={dashed, gray!25},
  axis lines*=left,
  tick align=outside,
  tick style={color=black!60},
  axis line style={color=black!60},
]
\addplot[fill=blue!55!white, draw=blue!75!black, line width=0.4pt]
  coordinates {(Q1,34.1)(Q2,38.2)(Q3,44.2)(Q4,43.6)};
\addplot[pattern=north east lines, pattern color=blue!75!black,
         draw=blue!75!black, line width=0.4pt]
  coordinates {(Q1,43.6)(Q2,41.9)(Q3,48.0)(Q4,51.8)};
\addplot[fill=orange!65!white, draw=orange!85!black, line width=0.4pt]
  coordinates {(Q1,17.3)(Q2,16.2)(Q3,18.8)(Q4,18.1)};
\addplot[pattern=north east lines, pattern color=orange!85!black,
         draw=orange!85!black, line width=0.4pt]
  coordinates {(Q1,19.3)(Q2,18.9)(Q3,22.7)(Q4,20.2)};
\end{axis}
\end{tikzpicture}
}

\vspace{2pt}

{\hfuzz=50pt
\begin{tikzpicture}[trim axis left, trim axis right]
\begin{axis}[
  ybar, bar width=0.10cm, ybar=1.5pt,
  width=0.95\columnwidth,
  height=4.2cm,
  enlarge x limits=0.18,
  ylabel={SPR (\%)},
  ylabel style={font=\footnotesize, yshift=-8pt},
  ymin=0, ymax=110,
  ytick={0,20,40,60,80,100},
  yticklabel style={font=\scriptsize},
  xtick=data,
  symbolic x coords={Q1,Q2,Q3,Q4},
  xticklabels={Easiest, Easy, Hard, Hardest},
  x tick label style={font=\scriptsize},
  xlabel={Task Difficulty},
  xlabel style={font=\footnotesize, yshift=-4pt},
  legend style={
    at={(0.5,-0.52)}, anchor=north,
    legend columns=2,
    font=\scriptsize,
    draw=none, fill=none,
    column sep=8pt,
    row sep=2pt,
  },
  legend cell align={left},
  legend image code/.code={
    \draw[#1, draw opacity=1]
      (0.00cm,0.00cm) rectangle (0.14cm,0.28cm);
    \draw[#1, draw opacity=1]
      (0.18cm,0.00cm) rectangle (0.32cm,0.18cm);
  },
  ymajorgrids=true,
  grid style={dashed, gray!25},
  axis lines*=left,
  tick align=outside,
  tick style={color=black!60},
  axis line style={color=black!60},
]
\addplot[fill=blue!55!white, draw=blue!75!black, line width=0.4pt]
  coordinates {(Q1,100.0)(Q2,80.1)(Q3,24.8)(Q4,0.8)};
\addlegendentry{GPT-5 Mini (Base)}
\addplot[pattern=north east lines, pattern color=blue!75!black,
         draw=blue!75!black, line width=0.4pt]
  coordinates {(Q1,94.9)(Q2,73.1)(Q3,32.6)(Q4,29.7)};
\addlegendentry{GPT-5 Mini (+CCE)}
\addplot[fill=orange!65!white, draw=orange!85!black, line width=0.4pt]
  coordinates {(Q1,99.9)(Q2,48.9)(Q3,1.2)(Q4,0.8)};
\addlegendentry{Qwen3-Coder (Base)}
\addplot[pattern=north east lines, pattern color=orange!85!black,
         draw=orange!85!black, line width=0.4pt]
  coordinates {(Q1,88.4)(Q2,41.2)(Q3,19.4)(Q4,21.9)};
\addlegendentry{Qwen3-Coder (+CCE)}
\end{axis}
\end{tikzpicture}
}

\caption{%
  Absolute Caller Coverage (top) and SPR (bottom) per difficulty quartile
  for the bash-only Agent vs.\ Agent\textsubscript{+CCE} across GPT-5 Mini and Qwen3-Coder 30B.
}
\label{fig:absolute_quartiles}
\vspace{-1.5em}
\end{figure}
\paragraph{Benchmark Scaling.}

\begin{table}[t]
\centering
\resizebox{\columnwidth}{!}{%
\begin{tabular}{lcccc}
\toprule
\textbf{Setting / Model}
  & \textbf{GPT-5 mini}
  & \textbf{Qwen3-Coder-30B}
  & \textbf{Devstral} \\
\midrule
Full-Context
  & 55.86\textsubscript{+1.02}
  & 38.36\textsubscript{+1.07}
  & 13.72\textcolor{red}{\textsubscript{$-$16.20}} \\
Full-Context + CoT
  & 55.28\textsubscript{$-$0.80}
  & 34.04\textcolor{red}{\textsubscript{$-$3.03}}
  & 14.20\textcolor{red}{\textsubscript{$-$11.34}} \\
Agent
  & 54.47\textcolor{blue}{\textsubscript{+5.08}}
  & 35.48\textcolor{red}{\textsubscript{$-$5.58}}
  & 33.74\textsubscript{+0.25} \\
Agent + CCE toolkit
  & 57.72\textsubscript{+1.58}
  & 42.60\textsubscript{$-$0.11}
  & 36.84\textcolor{red}{\textsubscript{$-$3.28}} \\
\bottomrule
\end{tabular}%
}
\caption{APR (\%) on large-scale benchmark instances. Subscripts denote change in percentage points from \recube.}
\label{tab:scale_spr}
\vspace{-1.5em}
\end{table} 


To understand how the increased repository-level context $\mathcal{C}_{f_{i}}$ affects reconstruction performance, we compare the results of three different models between \recube and \recubelarge.
Devstral collapses under full-context generation, with SPR dropping more than 10\%, primarily because large-scale instances (up to 338K) exceed its context capacity, where 39.13\% of cases fail due to direct context-length errors.
Using agentic frameworks mitigates much of the loss, as the input token dramatically decreases.
On the contrary, GPT-5 Mini exhibits better performance on \recubelarge, leveraging the expanded information through its context capacity, and improves even more in the agentic settings.

\paragraph{Cost Efficiency Analysis.}

We perform a cost-efficiency analysis of the evaluated agents (Table~\ref{tab:cost_efficiency}), presenting the per-instance costs, turns, and efficiency metrics.
Across both benchmarks, agents with our CCE toolkit demonstrate superior SPR-per-cost and SPR-per-turn compared to the baseline bash-only agents.
The performance gap widens on the large-scale benchmark, showing that our method is particularly effective in larger repository contexts.
These results indicate that the CCE toolkit effectively optimizes the agent's exploration trajectory, yielding navigation paths that are both efficient and concise.


\begin{table}[H]
\centering
\resizebox{\columnwidth}{!}{%
\begin{tabular}{llrrrr}
\toprule
\textbf{Scale} & \textbf{Method} & \textbf{Cost} & \textbf{Turns} & \textbf{SPR/Cost} & \textbf{SPR/Turn} \\
\midrule
\multirow{2}{*}{Original}
  & Agent          & \$ 0.1226 & 21.35 & 254.6 & 1.46 \\
  & Agent (Ours)   & \$ 0.1241 & 20.30 & \textbf{312.5} & \textbf{1.91} \\
\midrule
\multirow{2}{*}{Large}
  & Agent          & \$ 0.1410 & 25.66 & 221.4 & 1.22 \\
  & Agent (Ours)   & \$ 0.1166 & 19.81 & \textbf{332.6} & \textbf{1.96} \\
\bottomrule
\end{tabular}%
}
\caption{Cost-efficiency comparison for GPT-5-mini across two different agent settings and benchmark scales.}
\label{tab:cost_efficiency}
\vspace{-0.5em}
\end{table}

\section{Conclusion}

This work introduces \recube, a benchmark designed to evaluate how well LLMs leverage repository-level context during code generation.
We demonstrate that even state-of-the-art models struggle to reconstruct functionally correct, well-integrated code from repository context.
To better guide agents through large codebases, we propose the Caller-Centric Exploration (CCE) toolkit, which consistently achieves the best performance across all evaluated models and settings.
We hope \recube serves as a foundation for future research, encouraging the development of models and agents that go beyond surface-level pattern matching to effectively understand and navigate the structural dependencies of real-world codebases.

\section*{Limitations}
Because \recube is constructed from recent, popular repositories, it inherently favors prevalent codebase patterns.
Consequently, its domain scope is relatively narrow, predominantly focusing on LLM-related projects, and its implementation files are strictly restricted to Python.
Moreover, the high computational cost of the full \recube evaluation limited our study to a small subset of models, particularly in agent-based settings.
In future work, we plan to evaluate a broader range of models in both full-context and agent environments, leveraging additional frameworks like OpenHands~\cite{openhands} and SWE-agent~\cite{sweagent} to further assess the CCE toolkit's effectiveness.

\bibliography{custom}

@misc{codex,
      title={Evaluating Large Language Models Trained on Code}, 
      author={Mark Chen and Jerry Tworek and Heewoo Jun and Qiming Yuan and Henrique Ponde de Oliveira Pinto and Jared Kaplan and Harri Edwards and Yuri Burda and Nicholas Joseph and Greg Brockman and Alex Ray and Raul Puri and Gretchen Krueger and Michael Petrov and Heidy Khlaaf and Girish Sastry and Pamela Mishkin and Brooke Chan and Scott Gray and Nick Ryder and Mikhail Pavlov and Alethea Power and Lukasz Kaiser and Mohammad Bavarian and Clemens Winter and Philippe Tillet and Felipe Petroski Such and Dave Cummings and Matthias Plappert and Fotios Chantzis and Elizabeth Barnes and Ariel Herbert-Voss and William Hebgen Guss and Alex Nichol and Alex Paino and Nikolas Tezak and Jie Tang and Igor Babuschkin and Suchir Balaji and Shantanu Jain and William Saunders and Christopher Hesse and Andrew N. Carr and Jan Leike and Josh Achiam and Vedant Misra and Evan Morikawa and Alec Radford and Matthew Knight and Miles Brundage and Mira Murati and Katie Mayer and Peter Welinder and Bob McGrew and Dario Amodei and Sam McCandlish and Ilya Sutskever and Wojciech Zaremba},
      year={2021},
      eprint={2107.03374},
      archivePrefix={arXiv},
      primaryClass={cs.LG},
      url={https://arxiv.org/abs/2107.03374}, 
}

@inproceedings{intellicode,
    author = {Svyatkovskiy, Alexey and Deng, Shao Kun and Fu, Shengyu and Sundaresan, Neel},
    title = {IntelliCode compose: code generation using transformer},
    year = {2020},
    isbn = {9781450370431},
    publisher = {Association for Computing Machinery},
    address = {New York, NY, USA},
    url = {https://doi.org/10.1145/3368089.3417058},
    doi = {10.1145/3368089.3417058},
    booktitle = {Proceedings of the 28th ACM Joint Meeting on European Software Engineering Conference and Symposium on the Foundations of Software Engineering},
    pages = {1433–1443},
    numpages = {11},
    location = {Virtual Event, USA},
    series = {ESEC/FSE 2020}
}

@inproceedings{codexglue,
    title={Code{XGLUE}: A Machine Learning Benchmark Dataset for Code Understanding and Generation},
    author={Shuai Lu and Daya Guo and Shuo Ren and Junjie Huang and Alexey Svyatkovskiy and Ambrosio Blanco and Colin Clement and Dawn Drain and Daxin Jiang and Duyu Tang and Ge Li and Lidong Zhou and Linjun Shou and Long Zhou and Michele Tufano and MING GONG and Ming Zhou and Nan Duan and Neel Sundaresan and Shao Kun Deng and Shengyu Fu and Shujie LIU},
    booktitle={Thirty-fifth Conference on Neural Information Processing Systems Datasets and Benchmarks Track (Round 1)},
    year={2021},
    url={https://openreview.net/forum?id=6lE4dQXaUcb}
}

@misc{codescot,
      title={Understanding Chain-of-Thought Effectiveness in Code Generation: An Empirical and Information-Theoretic Analysis}, 
      author={Naizhu Jin and Zhong Li and Guang Yang and Tian Zhang and Qingkai Zeng},
      year={2025},
      eprint={2512.09679},
      archivePrefix={arXiv},
      primaryClass={cs.SE},
      url={https://arxiv.org/abs/2512.09679}, 
}

@inproceedings{codecot,
    title={Revisiting Chain-of-Thought in Code Generation: Do Language Models Need to Learn Reasoning before Coding?},
    author={Ren-Biao Liu and Anqi Li and Chaoding Yang and Hui Sun and Ming Li},
    booktitle={Forty-second International Conference on Machine Learning},
    year={2025},
    url={https://openreview.net/forum?id=wSZeQoJ1Vk}
}

@misc{moatless,
  author    = {Orjon, Albert},
  title     = {{Moatless Tools}},
  year      = {2024},
  url       = {https://github.com/aorwall/moatless-tools},
  note      = {GitHub repository}
}

@inproceedings{multiswebench,
    title={Multi-{SWE}-bench: A Multilingual Benchmark for Issue Resolving},
    author={Daoguang Zan and Zhirong Huang and Wei Liu and Hanwu Chen and Shulin Xin and Linhao Zhang and Qi Liu and Aoyan Li and Lu Chen and Xiaojian Zhong and Siyao Liu and Yongsheng Xiao and Liangqiang Chen and Yuyu Zhang and Jing Su and Tianyu Liu and RUI LONG and Ming Ding and liang xiang},
    booktitle={The Thirty-ninth Annual Conference on Neural Information Processing Systems Datasets and Benchmarks Track},
    year={2025},
    url={https://openreview.net/forum?id=MhBZzkz4h9}
}

@misc{swebenchverified,
  author    = {OpenAI},
  title     = {{Introducing SWE-bench Verified}},
  year      = {2024},
  url       = {https://openai.com/index/introducing-swe-bench-verified/},
  note      = {OpenAI Blog}
}

@misc{copilot,
  author    = {Dohmke, Thomas},
  title     = {{GitHub Copilot Workspace: Welcome to the Copilot-native developer environment}},
  year      = {2024},
  url       = {https://github.blog/news-insights/product-news/github-copilot-workspace/},
  note      = {GitHub Blog, April 29, 2024}
}

@inproceedings{openhands,
    title={OpenHands: An Open Platform for {AI} Software Developers as Generalist Agents},
    author={Xingyao Wang and Boxuan Li and Yufan Song and Frank F. Xu and Xiangru Tang and Mingchen Zhuge and Jiayi Pan and Yueqi Song and Bowen Li and Jaskirat Singh and Hoang H. Tran and Fuqiang Li and Ren Ma and Mingzhang Zheng and Bill Qian and Yanjun Shao and Niklas Muennighoff and Yizhe Zhang and Binyuan Hui and Junyang Lin and Robert Brennan and Hao Peng and Heng Ji and Graham Neubig},
    booktitle={The Thirteenth International Conference on Learning Representations},
    year={2025},
    url={https://openreview.net/forum?id=OJd3ayDDoF}
}

@article{agentless,
    author = {Xia, Chunqiu Steven and Deng, Yinlin and Dunn, Soren and Zhang, Lingming},
    title = {Demystifying LLM-Based Software Engineering Agents},
    year = {2025},
    issue_date = {July 2025},
    publisher = {Association for Computing Machinery},
    address = {New York, NY, USA},
    volume = {2},
    number = {FSE},
    url = {https://doi.org/10.1145/3715754},
    doi = {10.1145/3715754},
    journal = {Proc. ACM Softw. Eng.},
    month = jun,
    articleno = {FSE037},
    numpages = {24}
}

@inproceedings{swt-bench,
    title={{SWT}-Bench: Testing and Validating Real-World Bug-Fixes with Code Agents},
    author={Niels M{\"u}ndler and Mark Niklas Mueller and Jingxuan He and Martin Vechev},
    booktitle={The Thirty-eighth Annual Conference on Neural Information Processing Systems},
    year={2024},
    url={https://openreview.net/forum?id=9Y8zUO11EQ}
}

@inproceedings{infinitybench,
    title = "$\infty${B}ench: Extending Long Context Evaluation Beyond 100{K} Tokens",
    author = "Zhang, Xinrong  and
      Chen, Yingfa  and
      Hu, Shengding  and
      Xu, Zihang  and
      Chen, Junhao  and
      Hao, Moo  and
      Han, Xu  and
      Thai, Zhen  and
      Wang, Shuo  and
      Liu, Zhiyuan  and
      Sun, Maosong",
    editor = "Ku, Lun-Wei  and
      Martins, Andre  and
      Srikumar, Vivek",
    booktitle = "Proceedings of the 62nd Annual Meeting of the Association for Computational Linguistics (Volume 1: Long Papers)",
    month = aug,
    year = "2024",
    address = "Bangkok, Thailand",
    publisher = "Association for Computational Linguistics",
    url = "https://aclanthology.org/2024.acl-long.814/",
    doi = "10.18653/v1/2024.acl-long.814",
    pages = "15262--15277"
}

@misc{graphcodeagent,
      title={GraphCodeAgent: Dual Graph-Guided LLM Agent for Retrieval-Augmented Repo-Level Code Generation}, 
      author={Jia Li and Xianjie Shi and Kechi Zhang and Ge Li and Zhi Jin and Lei Li and Huangzhao Zhang and Jia Li and Fang Liu and Yuwei Zhang and Zhengwei Tao and Yihong Dong and Yuqi Zhu and Chongyang Tao},
      year={2025},
      eprint={2504.10046},
      archivePrefix={arXiv},
      primaryClass={cs.SE},
      url={https://arxiv.org/abs/2504.10046}, 
}

@inproceedings{codexgraph,
    title = "{C}odex{G}raph: Bridging Large Language Models and Code Repositories via Code Graph Databases",
    author = "Liu, Xiangyan  and
      Lan, Bo  and
      Hu, Zhiyuan  and
      Liu, Yang  and
      Zhang, Zhicheng  and
      Wang, Fei  and
      Shieh, Michael Qizhe  and
      Zhou, Wenmeng",
    editor = "Chiruzzo, Luis  and
      Ritter, Alan  and
      Wang, Lu",
    booktitle = "Proceedings of the 2025 Conference of the Nations of the Americas Chapter of the Association for Computational Linguistics: Human Language Technologies (Volume 1: Long Papers)",
    month = apr,
    year = "2025",
    address = "Albuquerque, New Mexico",
    publisher = "Association for Computational Linguistics",
    url = "https://aclanthology.org/2025.naacl-long.7/",
    doi = "10.18653/v1/2025.naacl-long.7",
    pages = "142--160",
    ISBN = "979-8-89176-189-6"
}

@inproceedings{draco,
    title = "Dataflow-Guided Retrieval Augmentation for Repository-Level Code Completion",
    author = "Cheng, Wei  and
      Wu, Yuhan  and
      Hu, Wei",
    editor = "Ku, Lun-Wei  and
      Martins, Andre  and
      Srikumar, Vivek",
    booktitle = "Proceedings of the 62nd Annual Meeting of the Association for Computational Linguistics (Volume 1: Long Papers)",
    month = aug,
    year = "2024",
    address = "Bangkok, Thailand",
    publisher = "Association for Computational Linguistics",
    url = "https://aclanthology.org/2024.acl-long.431/",
    doi = "10.18653/v1/2024.acl-long.431",
    pages = "7957--7977"
}

@misc{grace,
      title={GRACE: Graph-Guided Repository-Aware Code Completion through Hierarchical Code Fusion}, 
      author={Xingliang Wang and Baoyi Wang and Chen Zhi and Junxiao Han and Xinkui Zhao and Jianwei Yin and Shuiguang Deng},
      year={2025},
      eprint={2509.05980},
      archivePrefix={arXiv},
      primaryClass={cs.SE},
      url={https://arxiv.org/abs/2509.05980}, 
}

@inproceedings{graphcoder,
    author = {Liu, Wei and Yu, Ailun and Zan, Daoguang and Shen, Bo and Zhang, Wei and Zhao, Haiyan and Jin, Zhi and Wang, Qianxiang},
    title = {GraphCoder: Enhancing Repository-Level Code Completion via Coarse-to-fine Retrieval Based on Code Context Graph},
    year = {2024},
    isbn = {9798400712487},
    publisher = {Association for Computing Machinery},
    address = {New York, NY, USA},
    url = {https://doi.org/10.1145/3691620.3695054},
    doi = {10.1145/3691620.3695054},
    booktitle = {Proceedings of the 39th IEEE/ACM International Conference on Automated Software Engineering},
    pages = {570–581},
    numpages = {12},
    location = {Sacramento, CA, USA},
    series = {ASE '24}
}

@inproceedings{rpg,
    title={{RPG}: A Repository Planning Graph for Unified and Scalable Codebase Generation},
    author={Jane Luo and Xin Zhang and Steven Liu and Jie Wu and Yiming Huang and Yangyu Huang and Chengyu Yin and Ying Xin and Jianfeng Liu and Yuefeng Zhan and Hao Sun and Qi Chen and Scarlett Li and Mao Yang},
    booktitle={The Fourteenth International Conference on Learning Representations},
    year={2026},
    url={https://openreview.net/forum?id=VAQq3Y8tIF}
}

@article{alphatrans,
    author = {Ibrahimzada, Ali Reza and Ke, Kaiyao and Pawagi, Mrigank and Abid, Muhammad Salman and Pan, Rangeet and Sinha, Saurabh and Jabbarvand, Reyhaneh},
    title = {AlphaTrans: A Neuro-Symbolic Compositional Approach for Repository-Level Code Translation and Validation},
    year = {2025},
    issue_date = {July 2025},
    publisher = {Association for Computing Machinery},
    address = {New York, NY, USA},
    volume = {2},
    number = {FSE},
    url = {https://doi.org/10.1145/3729379},
    doi = {10.1145/3729379},
    journal = {Proc. ACM Softw. Eng.},
    month = jun,
    articleno = {FSE109},
    numpages = {23}
}

@inproceedings{codeplan,
    author = {Bairi, Ramakrishna and Sonwane, Atharv and Kanade, Aditya and C, Vageesh and Iyer, Arun and Parthasarathy, Suresh and Rajamani, Sriram and Ashok, B. and Shet, Shashank},
    year = {2023},
    month = {10},
    pages = {},
    title = {CodePlan: Repository-level Coding using LLMs and Planning},
    doi = {10.48550/arXiv.2309.12499}
}

@inproceedings{sweagent,
  title={{SWE}-agent: Agent-Computer Interfaces Enable Automated Software Engineering},
  author={John Yang and Carlos E Jimenez and Alexander Wettig and Kilian Lieret and Shunyu Yao and Karthik R Narasimhan and Ofir Press},
  booktitle={The Thirty-eighth Annual Conference on Neural Information Processing Systems},
  year={2024},
  url={https://arxiv.org/abs/2405.15793}
}

@inproceedings{apps,
 author = {Hendrycks, Dan and Basart, Steven and Kadavath, Saurav and Mazeika, Mantas and Arora, Akul and Guo, Ethan and Burns, Collin and Puranik, Samir and He, Horace and Song, Dawn and Steinhardt, Jacob},
 booktitle = {Proceedings of the Neural Information Processing Systems Track on Datasets and Benchmarks},
 editor = {J. Vanschoren and S. Yeung},
 pages = {},
 title = {Measuring Coding Challenge Competence With APPS},
 url = {https://datasets-benchmarks-proceedings.neurips.cc/paper_files/paper/2021/file/c24cd76e1ce41366a4bbe8a49b02a028-Paper-round2.pdf},
 volume = {1},
 year = {2021}
}

@article{zhang2025swebenchgoeslive,
  title={SWE-bench Goes Live!},
  author={Linghao Zhang and Shilin He and Chaoyun Zhang and Yu Kang and Bowen Li and Chengxing Xie and Junhao Wang and Maoquan Wang and Yufan Huang and Shengyu Fu and Elsie Nallipogu and Qingwei Lin and Yingnong Dang and Saravan Rajmohan and Dongmei Zhang},
  journal={arXiv preprint arXiv:2505.23419},
  year={2025}
}

@misc{anthropic2025claudecode,
  author    = {Anthropic},
  title     = {{Claude Code}},
  year      = {2025},
  url       = {https://www.anthropic.com/news/claude-opus-4-1},
  note      = {Agentic coding tool}
}

@misc{llm-kendal-tau,
      title={L-Eval: Instituting Standardized Evaluation for Long Context Language Models}, 
      author={Chenxin An and Shansan Gong and Ming Zhong and Xingjian Zhao and Mukai Li and Jun Zhang and Lingpeng Kong and Xipeng Qiu},
      year={2023},
      eprint={2307.11088},
      archivePrefix={arXiv},
      primaryClass={cs.CL},
      url={https://arxiv.org/abs/2307.11088}, 
}

@article{spearman,
  title={The proof and measurement of association between two things},
  author={Spearman, Charles},
  journal={The American Journal of Psychology},
  volume={15},
  number={1},
  pages={72--101},
  year={1904},
  publisher={JSTOR}
}

@article{kendal-tau,
    author = {KENDALL, M. G.},
    title = {A NEW MEASURE OF RANK CORRELATION},
    journal = {Biometrika},
    volume = {30},
    number = {1-2},
    pages = {81-93},
    year = {1938},
    month = {06},
    issn = {0006-3444},
    doi = {10.1093/biomet/30.1-2.81},
    url = {https://doi.org/10.1093/biomet/30.1-2.81},
    eprint = {https://academic.oup.com/biomet/article-pdf/30/1-2/81/423380/30-1-2-81.pdf},
}

@misc{code-pearson2,
      title={CodeJudge: Evaluating Code Generation with Large Language Models}, 
      author={Weixi Tong and Tianyi Zhang},
      year={2024},
      eprint={2410.02184},
      archivePrefix={arXiv},
      primaryClass={cs.LG},
      url={https://arxiv.org/abs/2410.02184}, 
}

@misc{code-pearson1,
      title={Human-Like Code Quality Evaluation through LLM-based Recursive Semantic Comprehension}, 
      author={Fangzhou Xu and Sai Zhang and Zhenchang Xing and Xiaowang Zhang and Yahong Han and Zhiyong Feng},
      year={2024},
      eprint={2412.00314},
      archivePrefix={arXiv},
      primaryClass={cs.SE},
      url={https://arxiv.org/abs/2412.00314}, 
}

@misc{cot,
      title={Chain-of-Thought Prompting Elicits Reasoning in Large Language Models}, 
      author={Jason Wei and Xuezhi Wang and Dale Schuurmans and Maarten Bosma and Brian Ichter and Fei Xia and Ed Chi and Quoc Le and Denny Zhou},
      year={2023},
      eprint={2201.11903},
      archivePrefix={arXiv},
      primaryClass={cs.CL},
      url={https://arxiv.org/abs/2201.11903}, 
}

@misc{cot-codegen,
      title={Structured Chain-of-Thought Prompting for Code Generation}, 
      author={Jia Li and Ge Li and Yongmin Li and Zhi Jin},
      year={2023},
      eprint={2305.06599},
      archivePrefix={arXiv},
      primaryClass={cs.SE},
      url={https://arxiv.org/abs/2305.06599}, 
}

@misc{claude3,
      title={The Claude 3 Model Family: A New Standard for Intelligence},
      author={Anthropic},
      year={2024},
      url={https://www.anthropic.com/news/claude-3-family},
}

@misc{gemini25,
      title={Gemini 2.5: Pushing the Frontier with Advanced Reasoning, Multimodality, Long Context, and Next Generation Agentic Capabilities},
      author={Gheorghe Comanici and Eric Bieber and Mike Schaekermann and others},
      year={2025},
      eprint={2507.06261},
      archivePrefix={arXiv},
      primaryClass={cs.CL},
      url={https://arxiv.org/abs/2507.06261},
}

@misc{gpt5,
      title={OpenAI GPT-5 System Card},
      author={Aaditya Singh and others},
      year={2025},
      eprint={2601.03267},
      archivePrefix={arXiv},
      primaryClass={cs.CL},
      url={https://arxiv.org/abs/2601.03267},
}

@misc{gptoss,
      title={gpt-oss-120b \& gpt-oss-20b Model Card}, 
      author={OpenAI},
      year={2025},
      eprint={2508.10925},
      archivePrefix={arXiv},
      primaryClass={cs.CL},
      url={https://arxiv.org/abs/2508.10925}, 
}

@misc{devstral,
      title={Devstral: Fine-tuning Language Models for Coding Agent Applications}, 
      author={Abhinav Rastogi and Adam Yang and Albert Q. Jiang and Alexander H. Liu and Alexandre Sablayrolles and Amélie Héliou and Amélie Martin and Anmol Agarwal and Andy Ehrenberg and Andy Lo and Antoine Roux and Arthur Darcet and Arthur Mensch and Baptiste Bout and Baptiste Rozière and Baudouin De Monicault and Chris Bamford and Christian Wallenwein and Christophe Renaudin and Clémence Lanfranchi and Clément Denoix and Corentin Barreau and Darius Dabert Devon Mizelle and Diego de las Casas and Elliot Chane-Sane and Emilien Fugier and Emma Bou Hanna and Gabrielle Berrada and Gauthier Delerce and Gauthier Guinet and Georgii Novikov and Graham Neubig and Guillaume Lample and Guillaume Martin and Himanshu Jaju and Jan Ludziejewski and Jason Rute and Jean-Malo Delignon and JeanHadrien Chabran and Joachim Studnia and Joep Barmentlo and Jonas Amar and Josselin Somerville Roberts and Julien Denize and Karan Saxena and Karmesh Yadav and Kartik Khandelwal and Khyathi Raghavi Chandu and Kush Jain and Lélio Renard Lavaud and Léonard Blier and Lingxiao Zhao and Louis Martin and Lucile Saulnier and Luyu Gao and Marie Pellat and Mathilde Guillaumin and Mathis Felardos and Matthieu Dinot and Maxime Darrin and Maximilian Augustin and Mickaël Seznec and Neha Gupta and Nikhil Raghuraman and Olivier Duchenne and Patricia Wang and Patrick von Platen and Patryk Saffer and Paul Jacob and Paul Wambergue and Paula Kurylowicz and Philomène Chagniot and Pierre Stock and Pravesh Agrawal and Rémi Delacourt and Roman Soletskyi and Romain Sauvestre and Sagar Vaze and Sanchit Gandhi and Sandeep Subramanian and Shashwat Dalal and Siddharth Gandhi and Soham Ghosh and Srijan Mishra and Sumukh Aithal and Szymon Antoniak and Teven Le Scao and Thibaut Lavril and Thibault Schueller and Thomas Foubert and Thomas Robert and Thomas Wang and Timothée Lacroix and Tom Bewley and Valeriia Nemychnikova and Victor Paltz and Virgile Richard and Wen-Ding Li and William Marshall and Xingyao Wang and Xuanyu Zhang and Yihan Wan and Yunhao Tang},
      year={2025},
      eprint={2509.25193},
      archivePrefix={arXiv},
      primaryClass={cs.SE},
      url={https://arxiv.org/abs/2509.25193}, 
}

@misc{qwen3,
      title={Qwen3 Technical Report}, 
      author={Qwen Team},
      year={2025},
      eprint={2505.09388},
      archivePrefix={arXiv},
      primaryClass={cs.CL},
      url={https://arxiv.org/abs/2505.09388}, 
}

@misc{livecodebench,
      title={LiveCodeBench: Holistic and Contamination Free Evaluation of Large Language Models for Code}, 
      author={Naman Jain and King Han and Alex Gu and Wen-Ding Li and Fanjia Yan and Tianjun Zhang and Sida Wang and Armando Solar-Lezama and Koushik Sen and Ion Stoica},
      year={2024},
      eprint={2403.07974},
      archivePrefix={arXiv},
      primaryClass={cs.SE},
      url={https://arxiv.org/abs/2403.07974}, 
}

@misc{repobench,
      title={RepoBench: Benchmarking Repository-Level Code Auto-Completion Systems}, 
      author={Tianyang Liu and Canwen Xu and Julian McAuley},
      year={2023},
      eprint={2306.03091},
      archivePrefix={arXiv},
      primaryClass={cs.CL},
      url={https://arxiv.org/abs/2306.03091}, 
}

@inproceedings{longbench,
    title = "{L}ong{B}ench v2: Towards Deeper Understanding and Reasoning on Realistic Long-context Multitasks",
    author = "Bai, Yushi  and
      Tu, Shangqing  and
      Zhang, Jiajie  and
      Peng, Hao  and
      Wang, Xiaozhi  and
      Lv, Xin  and
      Cao, Shulin  and
      Xu, Jiazheng  and
      Hou, Lei  and
      Dong, Yuxiao  and
      Tang, Jie  and
      Li, Juanzi",
    editor = "Che, Wanxiang  and
      Nabende, Joyce  and
      Shutova, Ekaterina  and
      Pilehvar, Mohammad Taher",
    booktitle = "Proceedings of the 63rd Annual Meeting of the Association for Computational Linguistics (Volume 1: Long Papers)",
    month = jul,
    year = "2025",
    address = "Vienna, Austria",
    publisher = "Association for Computational Linguistics",
    url = "https://aclanthology.org/2025.acl-long.183/",
    doi = "10.18653/v1/2025.acl-long.183",
    pages = "3639--3664",
    ISBN = "979-8-89176-251-0"
}

@misc{longcodebench,
      title={LongCodeBench: Evaluating Coding LLMs at 1M Context Windows}, 
      author={Stefano Rando and Luca Romani and Alessio Sampieri and Luca Franco and John Yang and Yuta Kyuragi and Fabio Galasso and Tatsunori Hashimoto},
      year={2025},
      eprint={2505.07897},
      archivePrefix={arXiv},
      primaryClass={cs.CL},
      url={https://arxiv.org/abs/2505.07897}, 
}

@misc{xml1,
  author = {{Anthropic}},
  title = {Use {XML} Tags to Structure Your Prompts},
  year = {2024},
  howpublished = {Claude API Documentation},
  url = {https://docs.anthropic.com/en/docs/build-with-claude/prompt-engineering/use-xml-tags},
  note = {Accessed: 2025}
}

@misc{xml2,
      title={XML Prompting as Grammar-Constrained Interaction: Fixed-Point Semantics, Convergence Guarantees, and Human-AI Protocols}, 
      author={Faruk Alpay and Taylan Alpay},
      year={2025},
      eprint={2509.08182},
      archivePrefix={arXiv},
      primaryClass={cs.PL},
      url={https://arxiv.org/abs/2509.08182}, 
}

@misc{locobench-agent,
      title={LoCoBench-Agent: An Interactive Benchmark for LLM Agents in Long-Context Software Engineering}, 
      author={Jielin Qiu and Zuxin Liu and Zhiwei Liu and Rithesh Murthy and Jianguo Zhang and Haolin Chen and Shiyu Wang and Ming Zhu and Liangwei Yang and Juntao Tan and Roshan Ram and Akshara Prabhakar and Tulika Awalgaonkar and Zixiang Chen and Zhepeng Cen and Cheng Qian and Shelby Heinecke and Weiran Yao and Silvio Savarese and Caiming Xiong and Huan Wang},
      year={2025},
      eprint={2511.13998},
      archivePrefix={arXiv},
      primaryClass={cs.SE},
      url={https://arxiv.org/abs/2511.13998}, 
}

@misc{locobench,
      title={LoCoBench: A Benchmark for Long-Context Large Language Models in Complex Software Engineering}, 
      author={Jielin Qiu and Zuxin Liu and Zhiwei Liu and Rithesh Murthy and Jianguo Zhang and Haolin Chen and Shiyu Wang and Ming Zhu and Liangwei Yang and Juntao Tan and Zhepeng Cen and Cheng Qian and Shelby Heinecke and Weiran Yao and Silvio Savarese and Caiming Xiong and Huan Wang},
      year={2025},
      eprint={2509.09614},
      archivePrefix={arXiv},
      primaryClass={cs.SE},
      url={https://arxiv.org/abs/2509.09614}, 
}

@misc{Commit0,
      title={Commit0: Library Generation from Scratch}, 
      author={Wenting Zhao and Nan Jiang and Celine Lee and Justin T Chiu and Claire Cardie and Matthias Gallé and Alexander M Rush},
      year={2024},
      eprint={2412.01769},
      archivePrefix={arXiv},
      primaryClass={cs.SE},
      url={https://arxiv.org/abs/2412.01769}, 
}

@inproceedings{locagent,
    title = "{L}oc{A}gent: Graph-Guided {LLM} Agents for Code Localization",
    author = "Chen, Zhaoling  and
      Tang, Robert  and
      Deng, Gangda  and
      Wu, Fang  and
      Wu, Jialong  and
      Jiang, Zhiwei  and
      Prasanna, Viktor  and
      Cohan, Arman  and
      Wang, Xingyao",
    editor = "Che, Wanxiang  and
      Nabende, Joyce  and
      Shutova, Ekaterina  and
      Pilehvar, Mohammad Taher",
    booktitle = "Proceedings of the 63rd Annual Meeting of the Association for Computational Linguistics (Volume 1: Long Papers)",
    month = jul,
    year = "2025",
    address = "Vienna, Austria",
    publisher = "Association for Computational Linguistics",
    url = "https://aclanthology.org/2025.acl-long.426/",
    doi = "10.18653/v1/2025.acl-long.426",
    pages = "8697--8727",
    ISBN = "979-8-89176-251-0"
}

@misc{swebench,
      title={SWE-bench: Can Language Models Resolve Real-World GitHub Issues?}, 
      author={Carlos E. Jimenez and John Yang and Alexander Wettig and Shunyu Yao and Kexin Pei and Ofir Press and Karthik Narasimhan},
      year={2024},
      eprint={2310.06770},
      archivePrefix={arXiv},
      primaryClass={cs.CL},
      url={https://arxiv.org/abs/2310.06770}, 
}

@misc{githubsearch,
      author = {{GitHub}},
      title = {{GitHub Code Search}},
      year = {2026},
      howpublished = {\url{https://github.com/search}},
      note = {Accessed: 2026-01-02}
    }

\appendix

\section{Repository and Functional Subset Details}
\label{appendix:functional_subsets}

\recube comprises 40 carefully curated functional subsets spanning six functional domains.
Table~\ref{tab:repo_details} provides an overview of each repository.

\begin{table*}[!ht]
\centering
\caption{Detailed breakdown of the 40 functional subsets in our benchmark. Each entry represents a specific functional subset of a larger repository, selected to target distinct development capabilities.}
\label{tab:repo_details}
\resizebox{\textwidth}{!}{
\begin{tabular}{lll}
\toprule
\textbf{Repository} & \textbf{Functionality} & \textbf{Key Components} \\
\midrule
OpenManus & Agent orchestration with sandbox execution & Base agent, ReAct pattern, tool execution \\
nanochat & GPT training infrastructure & SFT/RL training, data loading, checkpointing \\
DeepSeek-OCR & Vision-based OCR system & CLIP/SAM encoders, image/PDF processing \\
deer-flow & Workflow-based agent with RAG & Graph orchestration, web crawling, retrieval \\
openai-agents & Core runtime with guardrails & Agent execution, function schemas, prompts \\
openai-agents & Computer use \& file editing & Editor interface, computer control \\
openai-agents & Multi-agent handoffs \& visualization & Handoff filtering, workflow visualization \\
openai-agents & MCP server with persistent memory & MCP integration, SQLite sessions \\
openai-agents & Streaming chat completions & Stream handling, chat conversion \\
openai-agents & Tracing \& observability & Span data, trace processors, providers \\
openai-agents & Error handling \& computer control & Error tracing, guardrails \\
openai-agents & Real-time voice agent & Realtime API, audio formats, voice handoffs \\
openai-agents & Voice pipeline (STT/TTS) & Speech-to-text, text-to-speech workflow \\
serena & Semantic code editing & Symbol-level operations, code editor \\
serena & LSP server implementation & Protocol handler, IDE integration \\
blender-mcp & Blender MCP server & 3D modeling control, telemetry \\
VibeVoice & Streaming voice generation & Diffusion models, DPM solver \\
deepwiki-open & Multi-provider LLM API & OpenAI/DashScope/Ollama, embeddings \\
context-engineering & Multi-agent system with RAG & Research agent, document ingestion \\
context-engineering & Standalone RAG agent & CLI, embedding pipeline, DB utilities \\
DeepCode & Multi-interface coding assistant & CLI \& Streamlit UI, workflows \\
DeepCode & Code implementation agent & MCP tools, Git ops, code indexing \\
fastapi\_mcp & FastAPI MCP server & HTTP/SSE transports, Auth0 \\
RAG-Anything & Document processing RAG & Parser, processor, query system \\
memvid & Video chat with retrieval & Indexing, encoding, ffmpeg processing \\
trae-agent & Software engineering agent & Docker execution, bash/edit tools, CKG \\
trae-agent & Trae with MCP integration & MCP tools, code knowledge graph \\
langextract & Structured extraction with plugins & Multi-provider, Gemini/Ollama \\
strix & Core agent with LLM management & Memory compression, request queuing \\
strix & Tool runtime with Docker & File editing, web search, thinking tools \\
strix & Interactive tool managers & Python/terminal/proxy/graph sessions \\
strix & Browser automation \& CLI & Browser control, tab management \\
strix & Tool rendering system & UI component renderers registry \\
strix & Terminal user interface & Text-based UI, utilities \\
Wan2.1 & Video generation models & Text2video, image2video, VAE \\
Second-Me & Personalized LLM data generation & L0/L1 generators, DPO training \\
Second-Me & GGUF model utilities & Format conversion, quantization \\
Second-Me & Personalized AI API services & Chat/knowledge/role management \\
Second-Me & Complete AI application backend & Document processing, vector storage, MCP \\
Spark-TTS & Neural TTS with tokenization & BiCodec, FSQ, Triton runtime \\
\bottomrule
\end{tabular}
}
\end{table*}
\FloatBarrier

\section{Prompts and Specifications}

\subsection{Prompts for Baseline Settings}
\label{appendix:instance}
In this section, we provide the prompts for the four baseline settings we used for running \recube (\S\ref{ssec:baselines}, \S\ref{ssec:agent_cce}) on LLMs with and without the agentic framework. We also illustrate how each of the prompt is curated and list all the specification along the prompts.

\paragraph{Full-Context.}
We first describe the task formulation (e.g., objectives, input, and output specifications) and point out the exact file \hl{\texttt{target\_file}} to reconstruct.
We restrict the output format to a complete, runnable Python file so that we could automatically parse it and run evaluation by applying the patch to the instance.

We then append each instance, which is composed of three sections: \hl{\texttt{<readme>}}, \hl{\texttt{<dependencies>}}, and \hl{\texttt{<implementations>}}.
The \hl{\texttt{<readme>}} is a consecutive markdown files such as \hl{\texttt{README.md}} that are relevant to each instance.
The \hl{\texttt{<dependencies>}} is a description of each dependency from \hl{\texttt{requirements.txt}} or \hl{\texttt{Pyproject.toml}} in each collected repository.
The description for each dependency is a full project description from the Python Package Index\footnote{https://pypi.org} of its version.
The average token count (cl100k\_base encoding) of instances is 70,997 tokens and the maximum count is 110,874 tokens.

\nolinenumbers
\begin{lstlisting}[style=promptstyle]
Your task is to reconstruct the complete implementation of the file `{target_file}`.

The file has been provided with:
- All function signatures and docstrings intact
- All function bodies replaced with `raise NotImplementedError`
- All import statements removed

You need to:
1. Add all necessary import statements at the beginning of the file
2. Implement the complete function bodies based on:
   - The function signatures and docstrings
   - The README documentation
   - The dependencies and their descriptions
   - All other files in the codebase that are fully implemented
   - The overall project structure and design patterns

IMPORTANT OUTPUT FORMAT:
You must respond with ONLY the reconstructed file content wrapped in a code block as follows:

```python
# Your complete reconstructed file here
# Include all imports
# Include all function implementations
```

Do not include any explanations or additional text outside the code block.
The code block must contain the complete, runnable Python file with all imports and implementations.

Here is the project context:

{REPOSITORY LEVEL CONTEXT}
\end{lstlisting}
\linenumbers

\paragraph{Full-Context + CoT.}
As a baseline, we augment the full-context prompt with a structured chain-of-thought (CoT) reasoning template.
As shown above, the prompt instructs the model to explicitly reason through the repository's README, relevant dependencies, interacting implementations, and the target file's role before producing the final output, organized into four dedicated XML tags: \hl{\texttt{<readme>}}, \hl{\texttt{<dependencies>}}, \hl{\texttt{<implementations>}}, and \hl{\texttt{<file\_context>}}.
This structured decomposition aims to surface and organize repository-level context prior to generation.

\nolinenumbers
\begin{lstlisting}[style=promptstyle]
Your task is to reconstruct the complete implementation of the file `{target_file}`.

The file has been provided with:
- All function signatures and docstrings intact
- All function bodies replaced with `raise NotImplementedError`
- All import statements removed

You need to:
1. Add all necessary import statements at the beginning of the file
2. Implement the complete function bodies based on:
   - The function signatures and docstrings
   - The README documentation
   - The dependencies and their descriptions
   - All other files in the codebase that are fully implemented
   - The overall project structure and design patterns

INSTRUCTIONS:
- Keep your THOUGHT section concise but informative - focus on the most relevant details that directly impact implementation decisions
- In <readme></readme>: Extract 2-3 key points about project purpose and architecture
- In <dependencies></dependencies>: List only dependencies directly needed for `{target_file}`
- In <implementations></implementations>: Mention only the files/classes/functions that `{target_file}` directly uses or interacts with
- In <file_context></file_context>: Briefly describe `{target_file}`'s role and main implementation approach
- The FINAL_OUTPUT code block must contain the complete, runnable Python file with all imports and implementations

IMPORTANT OUTPUT FORMAT:
<format_example>
THOUGHT:
<readme>Agent framework for autonomous task execution. Architecture: modular agent system with pluggable tools, sandbox execution environment for safety, LLM-based reasoning with ReAct pattern.</readme>

<dependencies>openai for LLM API calls, docker for sandbox isolation, pydantic for schema validation in app/schema.py.</dependencies>

<implementations>Inherits from BaseAgent in app/agent/base.py which provides core agent loop. Uses Tool base class from app/tool/base.py for tool registry. Interacts with Sandbox from app/sandbox/core/sandbox.py for code execution. Imports Message and AgentConfig schemas from app/schema.py.</implementations>

<file_context>app/agent/react.py implements ReAct agent - combines reasoning and acting in iterative loops. Needs to: parse LLM responses for thought/action/observation pattern, manage tool execution through base class methods, handle error cases when actions fail, format conversation history for context window.</file_context>

FINAL_OUTPUT:
```python
# Complete reconstructed file for app/agent/react.py
# Include ALL necessary imports
# Include ALL complete function implementations (no pass statements or placeholder comments)

from typing import List, Optional, Dict, Any
import re
from app.agent.base import BaseAgent
from app.schema import Message, AgentConfig
from app.tool.base import Tool
from app.llm import LLMClient

class ReactAgent(BaseAgent):
    # ReAct agent that combines reasoning and acting

    def __init__(self, config: AgentConfig):
        super().__init__(config)
        self.llm_client = LLMClient(config.model_name)
        self.max_iterations = config.max_iterations or 10

    def execute(self, task: str) -> str:
        # Execute task using ReAct pattern with thought/action/observation loop
        history = []
        for i in range(self.max_iterations):
            response = self.llm_client.chat(self._format_prompt(task, history))
            thought, action, action_input = self._parse_response(response)
            history.append({{"thought": thought, "action": action}})

            if action == "finish":
                return action_input

            observation = self._execute_tool(action, action_input)
            history.append({{"observation": observation}})

        return "Max iterations reached"

    def _parse_response(self, response: str) -> tuple[str, str, str]:
        # Parse LLM response into thought, action, and input using regex
        thought_match = re.search(r"Thought: (.*?)\\n", response)
        action_match = re.search(r"Action: (.*?)\\n", response)
        input_match = re.search(r"Action Input: (.*?)\\n", response)

        thought = thought_match.group(1) if thought_match else ""
        action = action_match.group(1) if action_match else ""
        action_input = input_match.group(1) if input_match else ""

        return thought, action, action_input

    def _execute_tool(self, tool_name: str, tool_input: str) -> str:
        # Execute tool and return observation with error handling
        tool = self.tool_registry.get(tool_name)
        if not tool:
            return f"Error: Tool {{tool_name}} not found"

        try:
            return tool.run(tool_input)
        except Exception as e:
            return f"Error executing tool: {{str(e)}}"

    def _format_prompt(self, task: str, history: List[Dict]) -> str:
        # Format the prompt with task and conversation history
        prompt = f"Task: {{task}}\\n\\n"
        for entry in history:
            if "thought" in entry:
                prompt += f"Thought: {{entry['thought']}}\\n"
                prompt += f"Action: {{entry['action']}}\\n"
            if "observation" in entry:
                prompt += f"Observation: {{entry['observation']}}\\n"
        return prompt
```
</format_example>

Here is the project context:

{REPOSITORY LEVEL CONTEXT}
\end{lstlisting}
\linenumbers

\paragraph{Agent.}
We adopt the mini SWE-agent template and modify the task description to specify the exact file to reconstruct; all other configuration remains unchanged.

\nolinenumbers
\begin{lstlisting}[style=promptstyle]
agent:
  system_template: |
    ...

    <task_description>
    TASK: Implement the file `{{ target_file }}`

    The project has been set up at /workspace with:
    - All dependencies installed
    - All project files in place EXCEPT the target file
    - The target file `{{ target_file }}` exists but has:
      * All imports removed (file starts with just docstring and/or comments)
      * All function/class bodies replaced with `raise NotImplementedError()`

    Your objective:
    1. Add ALL necessary import statements to `{{ target_file }}`
    2. Implement ALL function and class bodies in `{{ target_file }}` based on:
       - Function signatures and docstrings in the file
       - The README documentation below
       - Other fully implemented files in the codebase
       - The overall project structure and design patterns

    IMPORTANT: You MUST implement EVERY function and class in the file. Leaving any `raise NotImplementedError()` is considered incomplete.
    </task_description>

    ...
  step_limit: 75

environment:
  cwd: "/workspace"
  timeout: 300
  env:
    PAGER: cat
    MANPAGER: cat
    LESS: -R
    PIP_PROGRESS_BAR: 'off'
    TQDM_DISABLE: '1'
  environment_class: docker

model:
  model_name: "gpt-5-mini"
  model_kwargs:
    drop_params: true
    temperature: 0.0
    max_completion_tokens: 128000
\end{lstlisting}
\linenumbers

\paragraph{Agent + CCE toolkit.}
We apply the CCE toolkit on top of the mini SWE-agent template to expose additional tools to the agent. Using these tools, the agent is able to explore the repositories in a more expected way. The tools are defined between \hl{\texttt{<available\_tools>}} and \hl{\texttt{</available\_tools>}}. We keep the rest of the configuration identical to the bash-only setting.

\nolinenumbers
\begin{lstlisting}[style=promptstyle]
agent:
  system_template: |
    ...
    
    <available_tools>
    ## Dependency Graph Tools (Caller-First Principle)

    These tools implement the caller-first principle to help agents explore systematically by providing repository-level understanding and caller patterns using the dependency graph before diving into implementation.

    **EVIDENCE**: Caller coverage correlates with significant improvement on moderately-hard tasks.

    Graph structure:
    - **Nodes**: directory, file, class, function
    - **Edges**: contains, imports, invokes, inherits

    ### 1. show_implementation_context (RECOMMENDED FIRST)
    Comprehensive repository understanding combining 4 graph-based perspectives.

    Usage:
    ```bash
    python /workspace/tools/graph_tools.py show_implementation_context --target {{ target_file }}
    ```

    Output (4 perspectives):
    1. **Caller Patterns** (incoming invoke/import edges) - Who uses your code and how
    2. **Inheritance** (inherit edges) - Parent classes and sibling implementations
    3. **Module Context** (directory structure) - Sibling files and common imports
    4. **Similar Files** (pattern matching) - Files with comparable structure

    ### 2. validate_code (MANDATORY)
    CRITICAL: Run this BEFORE submitting to prevent collection failures!
    Catches syntax errors that would cause 0% test pass rate.

    Usage:
    ```bash
    python /workspace/tools/graph_tools.py validate_code --file /workspace/{{ target_file }}
    ```

    Output: Syntax validation, implementation completeness, duplicate word detection.

    ### 3. Python Code Execution
    Query the graph programmatically for custom analysis.

    The graph_helper module provides easy-to-use functions:

    **Example 1: Find who uses a specific entity (RECOMMENDED)**
    ```bash
    python3 -c "
    import sys
    sys.path.insert(0, '/workspace/tools')
    from graph_helper import load_repo_graph, find_usages
    graph = load_repo_graph()
    # Find all files that use a specific class or function
    usages = find_usages(graph, '{{ target_file }}:YourClassName')
    print('Files using YourClassName:')
    for file in usages:
        print(f'  - {file}')
    "
    ```

    **Example 2: Search for entities**
    ```bash
    python3 -c "
    import sys
    sys.path.insert(0, '/workspace/tools')
    from graph_helper import quick_search
    results = quick_search('Config', 'BaseAgent')
    for name, nodes in results.items():
        print(f'{name}: {nodes}')
    "
    ```

    **Example 3: Get file content to see usage**
    ```bash
    python3 -c "
    import sys
    sys.path.insert(0, '/workspace/tools')
    from graph_helper import load_repo_graph, get_file_content
    graph = load_repo_graph()
    code = get_file_content(graph, 'app/caller_file.py')
    print(code)
    "
    ```

    **Example 4: List all classes in a file**
    ```bash
    python3 -c "
    import sys
    sys.path.insert(0, '/workspace/tools')
    from graph_helper import load_repo_graph, list_classes
    graph = load_repo_graph()
    classes = list_classes(graph, '{{ target_file }}')
    for cls in classes:
        print(cls)
    "
    ```

    **Available functions in graph_helper:**
    - `load_repo_graph()` - Load dependency graph
    - `search_entities(graph, names)` - Search for entities by name
    - `get_file_content(graph, file)` - Get full file source code
    - `get_entity_code(graph, node_id)` - Get class/function code
    - `list_files(graph)` - List all Python files
    - `list_classes(graph, file)` - List classes in file
    - `list_functions(graph, file)` - List functions in file
    - `find_usages(graph, entity_id)` - Find where entity is used (RECOMMENDED)
    - `get_imports_for_file(graph, file)` - Get import details
    - `quick_search(*names)` - Quick entity search
    - `get_dependencies(graph, file, depth)` - Get file dependencies (if needed)
    </available_tools>
    ...
\end{lstlisting}
\linenumbers

\subsection{Prompts for Generating Test Cases}
\label{appendix:test-cases}
In this section, we provide the prompts for generating test cases. As mentioned in \S\ref{sec:test-case}, we used claude code (Claude Opus 4.1) and manual verfication to synthesize the unit tests for each instance. For each functional subsets, we prompt the following text to the claude code. The specific prompt is illustrate in Appendix~\ref{appendix:test-cases}.

\nolinenumbers
\begin{lstlisting}[style=promptstyle]
# Test Case Generation and Validation Task

## Overview
Your task is to create comprehensive test cases for repository instances identified by `repo_id N`. Each instance corresponds to a file listed in `selected_files` under `PATH/TO/DIR/target_verified.json` where the `repo_id` is N.

## Task 1: Create Test Files

### Requirements
1. **Location**: Create test files under `PATH/TO/DIR/tests/{repo_id}/`
2. **Naming Convention**: Create exactly one test file per instance
   - Format: `{file_name}_test.py`
   - Example: For `app/agent/base.py` -> create `app_agent_base_test.py`
   - Replace path separators (`/`) with underscores (`_`)

### Test Case Standards
Your test cases must be **usage-aware unit tests** generated from real codebase interactions.

**CRITICAL REQUIREMENT - Test Difficulty and Discrimination:**
- **Tests MUST be challenging enough to distinguish correct from incorrect implementations**
- Tests should **pass on gold (original) implementation** but **fail on incorrect model-generated patches**
- Design tests that are **sensitive to implementation details**:
  - Test edge cases, boundary conditions, and corner cases
  - Verify exact return values, types, and data structures
  - Test error handling and exception cases
  - Validate state changes and side effects
  - Check integration between components
- **Avoid trivial tests** that only check basic functionality - they won't catch subtle bugs
- Each test should serve as a **discriminator** between correct and incorrect code
- If a model generates slightly wrong code, your tests should catch it

1. **Analysis Phase - Understanding the Repository Context**

   The complete repository code is available in `PATH/TO/DIR/instructions/{repo_id}.txt`:

   **File Structure:**
   - `<readme>...</readme>`: Repository documentation
   - `<dependencies>...</dependencies>`: Required packages
   - `<implementations>...</implementations>`: **All repository source code files**

   **Within `<implementations>` section:**
   - Each file is marked with: `## path/to/file.py`
   - Followed by a code block: ` ```python ... ``` `
   - Example from `0.txt`:
     ```
     ## app/agent/base.py
     ```python
     [full source code here]
     ```

     ## app/schema.py
     ```python
     [full source code here]
     ```
     ```

   **Analysis Tasks:**
   - [ ] Read the instruction file `PATH/TO/DIR/instructions/{repo_id}.txt`
   - [ ] Parse the `<implementations>` section to access all repository files
   - [ ] For the target file you're testing, analyze how its functions/classes are called:
     - Search for imports of the target file across other files
     - Find function/class invocations with example parameters
     - Identify usage patterns in different contexts (internal and cross-module)
   - [ ] Extract real usage contexts with file:line references
   - [ ] Identify typical call patterns, edge cases, and error conditions

2. **Test Generation Phase**
   - [ ] Generate tests that cover typical call patterns
   - [ ] **CRITICAL**: Include comprehensive edge case coverage that will fail on incorrect implementations
   - [ ] Test boundary conditions (empty inputs, None values, maximum/minimum values)
   - [ ] Test error conditions and exception handling with specific exception types
   - [ ] Verify exact outputs, not just "something was returned"
   - [ ] Test state changes, side effects, and interactions between methods
   - [ ] Follow the exact format demonstrated in example test files (e.g., `PATH/TO/DIR/tests/0/app_schema_test.py`)
   - [ ] **IMPORTANT**: Ensure all tests are CPU-runnable even for GPU-related code
     - Use mocking for GPU operations (e.g., `torch.cuda`, GPU kernels)
     - Test mock functionality and logic flow without requiring actual GPU hardware
     - Avoid dependencies on CUDA/GPU availability in test assertions

3. **Required Format Elements**
   - [ ] Proper docstrings for each test class and method
   - [ ] Test strategy documentation
   - [ ] Real usage references in comments (file:line format)
   - [ ] Clear test class organization
   - [ ] Descriptive test method names

### Reference Example
See `PATH/TO/DIR/tests/0/app_schema_test.py` for:
- Proper docstring format with test metadata header
- Test organization patterns (test classes grouping related functionality)
- Usage reference annotations (e.g., "Real usage: app/llm.py:1439")
- Test naming conventions (descriptive method names)
- How to document test strategy in docstrings

**Example of usage-aware test from `app_schema_test.py`:**
```python
def test_user_message_with_image(self):
    """
    Test user_message accepts base64_image.

    Real usage: app/agent/browser.py:2388 - Message.user_message(content=..., base64_image=...)
    """
    msg = Message.user_message("Check this", base64_image="abc123")

    assert msg.role == Role.USER
    assert msg.content == "Check this"
    assert msg.base64_image == "abc123"
```

Notice how the docstring references the actual line where this pattern is used in the repository (found by analyzing `app/agent/browser.py` in the `<implementations>` section).

### Examples: Good vs Bad Tests

**BAD TEST (Too trivial - won't catch bugs):**
```python
def test_add_message(self):
    """Test that add_message works."""
    memory = Memory()
    msg = Message.user_message("Hello")
    memory.add_message(msg)
    assert len(memory.messages) > 0  # Too weak!
```

**GOOD TEST (Specific and discriminating):**
```python
def test_add_message_enforces_max_messages(self):
    """
    Test add_message enforces max_messages limit.

    Real usage: app/schema.py:2077-2078
    if len(self.messages) > self.max_messages:
        self.messages = self.messages[-self.max_messages :]
    """
    memory = Memory(max_messages=3)

    # Add 5 messages
    for i in range(5):
        memory.add_message(Message.user_message(f"Message {i}"))

    # Should only keep last 3
    assert len(memory.messages) == 3
    assert memory.messages[0].content == "Message 2"
    assert memory.messages[2].content == "Message 4"
```

**Why the good test is better:**
- Tests specific behavior (max_messages enforcement)
- Verifies exact values, not just existence
- Would fail if implementation incorrectly kept first 3 instead of last 3
- Would fail if implementation didn't enforce limit at all
- Would fail if off-by-one error in slicing

### Validation Strategy

When validating your tests in Docker (Task 2), you will verify that:

1. **All tests PASS on gold implementation** - This proves your tests are correct
2. **Tests are discriminating** - To verify this, you can:
   - Mentally walk through: "What if the model makes these common mistakes?"
   - Consider: Empty list instead of None, wrong slice direction, missing validation, etc.
   - Your tests should catch these mistakes

**Remember**: The goal is NOT just to have passing tests. The goal is to have tests that will **fail when given incorrect code**. Think adversarially - what could a model get wrong, and does your test catch it?

## Task 2: Docker Validation and Fixing

### Step 1: Understand the Environment
- [ ] Review `PATH/TO/DIR/tests/docker/build_test_images.py` to understand:
  - Docker image structure
  - Testbed directory layout (`/testbed`)
  - How dependencies are installed
  - How files are copied into the image

### Step 2: Build or Verify Docker Image
**Image Details:**
- Image name: `DOCKER_IMAGE_NAME`
- Tag: `{repo_id}`
- Full image reference: `DOCKER_IMAGE_NAME:{repo_id}`

**Build Process:**
```bash
# If image doesn't exist, build it:
python PATH/TO/DIR/tests/docker/build_test_images.py --repo-id {repo_id}

# The image will contain:
# - Base Python environment
# - All dependencies from PATH/TO/DIR/instructions/{repo_id}.txt
# - Implementation files at /testbed/ (from selected_files)
# - Test files at /testbed/tests/
# - README.md
```

**What Gets Tested:**
- **During validation** (test_image.sh): Tests run against implementation files at `/testbed/`
- **During model evaluation**: Model patches replace files at `/testbed/`, then tests run
- Tests import directly from `/testbed/` via `PYTHONPATH=/testbed`

- [ ] Check if image `DOCKER_IMAGE_NAME:{repo_id}` exists
- [ ] If not, build using the script above
- [ ] Verify build completes successfully
- [ ] Push image to local Docker hub

### Step 3: Run Tests in Docker
```bash
# Run all tests for repo_id N:
PATH/TO/DIR/tests/docker/test_image.sh {repo_id}

# Or manually:
docker run --rm DOCKER_IMAGE_NAME:{repo_id} pytest tests/ -v --tb=short
```

- [ ] Execute test script
- [ ] Review all test results
- [ ] Document any failures

### Step 4: Fix Issues (Iterative Process)

#### Option 1: Fix Test Cases
Use when tests are incorrect or unstable:
- [ ] Identify problematic test cases
- [ ] Analyze why they fail (logic error, wrong assumptions, etc.)
- [ ] Modify test files under `PATH/TO/DIR/tests/{repo_id}/`
- [ ] **REQUIRED**: Rebuild image to include updated tests:
  ```bash
  python PATH/TO/DIR/tests/docker/build_test_images.py --repo-id {repo_id}
  ```
- [ ] **REQUIRED**: Verify the image was rebuilt (check timestamps/image ID)
- [ ] Re-run tests to verify fixes

#### Option 2: Fix Environment
Use when missing files/dependencies cause failures:

**Common Issues:**
- Missing implementation files not included in the initial `selected_files` list
- Missing internal dependencies/modules that are imported but not in `selected_files`
- Missing configuration files (e.g., `__init__.py`, config files)
- Missing supporting modules required by the code under test

**Resolution Steps:**
1. [ ] Identify missing files from error messages
2. [ ] Locate repository URL in `PATH/TO/DIR/reference/target_verified.json` (field: `repo_url` with matching `repo_id`)
3. [ ] Fetch missing files from the repository URL (download them to appropriate locations)
4. [ ] Modify `PATH/TO/DIR/tests/docker/build_test_images.py` to:
   - Add missing files to the build process (e.g., download files, copy from new locations)
   - Update file copying logic
   - Add any special installation steps or dependencies
   - **CRITICAL**: Ensure changes are permanent in the build script, not one-time fixes
5. [ ] **REQUIRED**: Rebuild the image completely:
   ```bash
   python PATH/TO/DIR/tests/docker/build_test_images.py --repo-id {repo_id}
   ```
6. [ ] **REQUIRED**: Verify the new image has replaced the old one:
   ```bash
   docker images DOCKER_IMAGE_NAME:{repo_id}
   # Check the CREATED timestamp - should be recent
   # Note the new IMAGE ID
   ```
7. [ ] **REQUIRED**: The rebuilt image is automatically the new version (Docker overwrites the same tag)
8. [ ] Re-run tests using the updated image to verify all fixes are included

### Step 5: Final Validation and Image Confirmation

**Before marking complete, verify:**
- [ ] All tests pass in Docker environment
- [ ] No pytest failures or errors
- [ ] Test coverage is comprehensive
- [ ] All edge cases are handled
- [ ] Documentation is complete
- [ ] **CRITICAL**: The final Docker image `DOCKER_IMAGE_NAME:{repo_id}` contains ALL modifications:
  - [ ] All test file updates are included
  - [ ] All environment fixes (missing files, dependencies) are included
  - [ ] Image was rebuilt after the last modification
  - [ ] Running tests on this image produces passing results

**Final Image Verification:**
```bash
# Confirm the image exists and is recent
docker images DOCKER_IMAGE_NAME:{repo_id}

# Run final test to confirm everything passes
PATH/TO/DIR/tests/docker/test_image.sh {repo_id}

# Expected: All tests pass, exit code 0
```

**Important**: The Docker image with tag `DOCKER_IMAGE_NAME:{repo_id}` IS the final version. Each rebuild automatically replaces the previous image with the same tag. No separate "push" step is needed for local images.

## Important Notes

1. **Environment Structure**: The Docker testbed uses `/testbed` as the working directory with `PYTHONPATH=/testbed`
   - Implementation files: `/testbed/{file_path}` (e.g., `/testbed/nanochat/gpt.py`)
   - Test files: `/testbed/tests/` (e.g., `/testbed/tests/nanochat_gpt_test.py`)

2. **Test Import Pattern**:
   - Import directly from `/testbed/` using the `PYTHONPATH=/testbed` environment variable
   - Example: `from app.schema import Message` -> imports from `/testbed/app/schema.py`
   - Example: `from nanochat.gpt import GPT` -> imports from `/testbed/nanochat/gpt.py`
   - **This is the standard pattern** - test files in `/testbed/tests/` import from implementation files at `/testbed/`

3. **Repository Completeness**:
   - The Docker environment initially contains only the **selected files** from `selected_files` in `target_verified.json`
   - **However**, you MUST add any additional files from the repository if they are required for tests to run
   - Fetch missing files from the `repo_url` and modify `build_test_images.py` to include them
   - The final environment should contain all necessary files for tests to execute successfully

4. **Iterative Process**: You may need multiple iterations of fixing and testing

5. **Image Updates - CRITICAL FOR FINAL VERSION**:
   - **ALWAYS** rebuild the image after ANY modification (test files OR environment files)
   - The rebuild process automatically replaces the old image with the same tag
   - The image tag `DOCKER_IMAGE_NAME:{repo_id}` always points to the most recent build
   - **Do NOT skip rebuilding** - the image will contain stale code otherwise
   - After ALL fixes are complete, verify the final image passes all tests
   - The last successfully built image IS the final deliverable version

6. **CPU-Only Testing**: All tests must run on CPU without requiring GPU hardware
   - Mock GPU operations using libraries like `unittest.mock` or `pytest-mock`
   - For PyTorch GPU code, mock `torch.cuda.is_available()` to return `False` or mock device placement
   - Test the logic and control flow, not actual GPU computation
   - Use `conftest.py` for shared mocking fixtures if needed across multiple test files

## Getting Help
If you encounter any of the following, please ask for assistance:
- Uncertainty about test case design
- Inability to locate missing dependencies
- Persistent test failures after multiple fix attempts
- Unclear error messages
- Questions about the Docker environment structure

## Checklist Summary

### Test Creation
- [ ] All test files created with correct naming
- [ ] Tests follow example format from reference files
- [ ] Usage-aware test cases generated
- [ ] Proper docstrings and test strategies documented

### Docker Validation
- [ ] Docker image built/verified initially
- [ ] Initial test run completed
- [ ] All failures analyzed
- [ ] Fixes applied (tests or environment)
- [ ] **CRITICAL**: Image rebuilt after EVERY change (test or environment)
- [ ] Image rebuild verified (check timestamp/IMAGE ID changed)
- [ ] Final validation: all tests pass on the final rebuilt image
- [ ] Confirmed final image `DOCKER_IMAGE_NAME:{repo_id}` contains all modifications

### Quality Assurance
- [ ] **CRITICAL**: All tests pass on gold (original) implementation
- [ ] **CRITICAL**: Tests are difficult enough to catch incorrect implementations
- [ ] Code coverage is comprehensive (not just adequate)
- [ ] Edge cases are thoroughly tested with specific assertions
- [ ] Boundary conditions are tested (empty, None, min/max values)
- [ ] Error conditions are handled with specific exception type checks
- [ ] Tests verify exact values, types, and data structures (not just "truthy" checks)
- [ ] Tests are stable and reproducible
- [ ] Documentation is clear and complete
- [ ] All tests run successfully on CPU (no GPU required)
- [ ] GPU-related code is properly mocked
\end{lstlisting}
\linenumbers

\section{Docstring Completeness Categories}
\label{appendix:docstrings}

The functions in our benchmark are categorized by docstring completeness into four levels, reflecting the varying degrees of guidance available to models during reconstruction.

\paragraph{Detailed Docstrings (10.9\%)} include comprehensive sections such as \texttt{Args}, \texttt{Returns}, \texttt{Examples}, and \texttt{Raises}, providing complete API documentation that fully specifies the function's interface and expected behavior. Example function:

\begin{lstlisting}[style=promptstyle, language=Python]
# Ex) app/llm.py - LLM.format_messages():
def format_messages(messages, supports_images):
    """Format messages for LLM processing.

    Args:
        messages: List of messages that can be either dict or Message objects
        supports_images: Flag indicating if the target model supports image inputs

    Returns:
        List[dict]: List of formatted messages in OpenAI format

    Raises:
        ValueError: If messages are invalid or missing required fields
        TypeError: If unsupported message types are provided

    Examples:
        >>> msgs = [
        ...     Message.system_message("You are a helpful assistant"),
        ...     {"role": "user", "content": "Hello"},
        ...     Message.user_message("How are you?")
        ... ]
        >>> formatted = LLM.format_messages(msgs)
        # function body
\end{lstlisting}

\paragraph{Args/Returns Docstrings (7.2\%)} document parameters and return values but omit usage examples and exception details, requiring models to infer edge cases and usage patterns. Example function:

\begin{lstlisting}[style=promptstyle, language=Python]
# Ex) app/llm.py - LLM.ask():
def ask(messages, system_msgs, stream, temperature):
    """Generate a response from the LLM.
  
    Args:
        messages: List of conversation messages
        system_msgs: Optional system messages to prepend
        stream (bool): Whether to stream the response
        temperature (float): Sampling temperature for the response

    Returns:
        str: The generated response

    Raises:
        TokenLimitExceeded: If token limits are exceeded
        ValueError: If messages are invalid or response is empty
        OpenAIError: If API call fails after retries
    """
    # function body
\end{lstlisting}

\paragraph{Signature-only Docstrings (27.5\%)} provide only a one-line description of what the function does without parameter or return documentation, leaving the implementation details largely unspecified. Example function:

\begin{lstlisting}[style=promptstyle, language=Python]
# Ex) app/schema.py - Message.system_message()
def system_message(cls, content: str):
    """Create a system message"""
\end{lstlisting}

\paragraph{No Docstrings (54.4\%)} have no documentation whatsoever, requiring models to infer behavior entirely from the function signature, variable names, and surrounding implementation context. Example function:

\begin{lstlisting}[style=promptstyle, language=Python]
# Ex) From app/bedrock.py - BedrockClient.__init__():
def __init__(self):
    # function body
\end{lstlisting}
\FloatBarrier

\section{Experiment Configurations}
\label{appendix:config}

\subsection{Bare-Bone LLM Configuration}

For static (full-context) settings, we configure LLMs as Tables~\ref{tab:config-open-qwen}, \ref{tab:config-open-mistral}, and \ref{tab:config-open-oss}.

\begin{table}[htb]
\centering
\small
\begin{tabular}{ll}
\toprule
\textbf{Parameter} & \textbf{Value} \\
\midrule
Temperature & 0.0 \\
\multirow{2}{*}{Max Tokens} & 8,192 (Full-Context) \\
                                        & 16,384 (Full-Context + CoT)\\
Timeout (s) & 300 \\
Inference Backend & vLLM == 0.17.0 \\
\texttt{enable\_auto\_tool\_choice} & False (Default) \\
\texttt{tool\_call\_parser} & None (Default) \\
\texttt{tokenizer\_mode} & auto (Default) \\
\texttt{max-model-len} & 262,144 (Default) \\
\bottomrule
\end{tabular}
\caption{VLLM configuration for Qwen/Qwen3-Coder-30B-A3B-Instruct.}
\label{tab:config-open-qwen}
\end{table}

\begin{table}[htb]
\centering
\small
\begin{tabular}{ll}
\toprule
\textbf{Parameter} & \textbf{Value} \\
\midrule
Temperature & 0.0 \\
\multirow{2}{*}{Max Tokens} & 8,192 (Full-Context) \\
                                        & 16,384 (Full-Context + CoT)\\
Timeout (s) & 300 \\
Inference Backend & vLLM == 0.17.0 \\
\texttt{config-format} & mistral \\
\texttt{load-format} & mistral \\
\texttt{tokenizer\_mode} & mistral \\
\texttt{max-model-len} & 131,072 (Default) \\
\bottomrule
\end{tabular}
\caption{VLLM configuration for mistralai/Devstral-Small-2507.}
\label{tab:config-open-mistral}
\end{table}

\begin{table}[htb]
\centering
\small
\begin{tabular}{ll}
\toprule
\textbf{Parameter} & \textbf{Value} \\
\midrule
Temperature & 0.0 \\
\multirow{2}{*}{Max Tokens} & 8,192 (Full-Context) \\
                                        & 16,384 (Full-Context + CoT)\\
Timeout (s) & 300 \\
Inference Backend & vLLM == 0.17.0 \\
\texttt{max-model-len} & 131,072 (Default) \\
\bottomrule
\end{tabular}
\caption{VLLM configuration for openai/gpt-oss-20b.}
\label{tab:config-open-oss}
\end{table}

\subsection{Agent Configuration}

We configure mini-SWE-agent (version 1.17.4) with the key parameters summarized in 
Tables~\ref{tab:config-proprietary} and~\ref{tab:config-open}. We use \hl{\texttt{max\_completion\_tokens}} instead of \hl{\texttt{max\_tokens}} for proprietary models since GPT-5 and GPT-5 Mini are reasoning models that internally generate extended chain-of-thought traces prior to producing a final response.

\begin{table}[htb]
\centering
\small
\begin{tabular}{ll}
\toprule
\textbf{Parameter} & \textbf{Value} \\
\midrule
Max Completion Tokens & 128,000 \\
Step Limit & 75 \\
Environment & Docker \\
Timeout (s) & 300 \\
\bottomrule
\end{tabular}
\caption{mini-SWE-agent configuration for proprietary models (GPT-5, GPT-5 Mini).}
\label{tab:config-proprietary}
\end{table}

\begin{table}[htb]
\centering
\small
\begin{tabular}{ll}
\toprule
\textbf{Parameter} & \textbf{Value} \\
\midrule
Temperature & 0.0 \\
Max Completion Tokens & 8,192 \\
Step Limit & 75 \\
Environment & Apptainer \\
Timeout (s) & 300 \\
Inference Backend & vLLM == 0.17.0 \\
GPU & NVIDIA H200 \\
\bottomrule
\end{tabular}
\caption{mini-SWE-agent configuration for open models (Qwen3-Coder 30B, Devstral Small 2507).}
\label{tab:config-open}
\end{table}

\subsection{GPT-OSS 20B on Agentic Frameworks}

Since GPT-OSS 20B is exclusively trained on OpenAI's proprietary Harmony response format~\cite{gptoss}, it is fundamentally incompatible with mini-SWE-agent, which relies on standard chat templates for agentic interaction. As a result, we exclude GPT-OSS 20B from the Agent and Agent + CCE toolkit settings in Table~\ref{tab:main_results}, and report its agent-setting results separately in Table~\ref{tab:oss-agent}.
To quantify this incompatibility, we analyze the agent trajectories and find that GPT-OSS 20B produces 25.60 and 25.28 format errors per instance under the base agent and Agent + CCE settings, respectively.
Notably, despite CCE being enabled, GPT-OSS 20B issues zero calls to any CCE tool, indicating that the model fails to parse the agentic scaffolding entirely and cannot leverage repository-level context guidance.

\begin{table}[t]
\centering
\scriptsize
\setlength{\tabcolsep}{3pt}
\begin{tabular}{l rr rr rr}
\toprule
& \multicolumn{2}{c}{\textbf{Overall}} 
& \multicolumn{2}{c}{\textbf{Ext. TCs}} 
& \multicolumn{2}{c}{\textbf{Int. TCs}} \\
\cmidrule(lr){2-3} \cmidrule(lr){4-5} \cmidrule(lr){6-7}
\textbf{Setting} 
& \textbf{SPR} & \textbf{APR} 
& \textbf{SPR} & \textbf{APR} 
& \textbf{SPR} & \textbf{APR} \\
\midrule
Agent           & 25.06 & 36.83 & 22.19 & 32.61 & 30.63 & 39.11 \\
Agent + CCE     & 21.01 & 30.49 & 16.40 & 23.95 & 24.32 & 33.20 \\
\bottomrule
\end{tabular}
\caption{GPT-OSS 20B agent results reported separately from
Table~\ref{tab:main_results} due to its incompatibility with
mini-SWE-agent, which stems from its exclusive training on
OpenAI's Harmony response format~\citet{gptoss}.}
\label{tab:oss-agent}
\end{table}

\section{CCE Toolkit Specifications}
\label{appendix:caller-first-principle}

In this section, we illustrate each category of the CCE toolkit with a detailed description and an example output.

\subsection{Caller Patterns}

By analyzing incoming invokes and imports edges, this feature provides a comprehensive overview of how external files utilize specific classes, functions, and methods within the target file.
It prioritizes a caller-centric perspective, mapping entity-level interactions and frequency to clarify usage patterns.
Example output:

\nolinenumbers
\begin{lstlisting}[style=promptstyle]
[1/4] CALLER PATTERNS - How your code is used
---
Found 8 file(s) that use your code:
use-cases/agent-factory-with-subagents/agents/rag_agent/cli.py
   invokes: DocumentIngestionPipeline.initialize (4x)
   invokes: main (4x)
use-cases/agent-factory-with-subagents/agents/rag_agent/dependencies.py
   invokes: DocumentIngestionPipeline.close (5x)
   invokes: DocumentIngestionPipeline.initialize (5x)
use-cases/agent-factory-with-subagents/agents/rag_agent/tests/conftest.py
   invokes: DocumentIngestionPipeline.close (15x)
use-cases/agent-factory-with-subagents/agents/rag_agent/tests/test_dependencies.py
   invokes: DocumentIngestionPipeline.initialize (35x)
use-cases/agent-factory-with-subagents/agents/rag_agent/tests/test_integration.py
   invokes: DocumentIngestionPipeline.initialize (19x)
use-cases/agent-factory-with-subagents/agents/rag_agent/tests/test_requirements.py
   invokes: DocumentIngestionPipeline.initialize (33x)
use-cases/agent-factory-with-subagents/agents/rag_agent/utils/db_utils.py
   invokes: DocumentIngestionPipeline.close (10x)
   invokes: DocumentIngestionPipeline.initialize (10x)
\end{lstlisting}
\linenumbers


\subsection{Inheritance}

This feature maps class hierarchies by analyzing inherits edges to identify parent classes and their alternative implementations. By traversing these relationships, it visualizes the target’s lineage and uncovers sibling classes, helping LLMs understand shared behaviors and inheritance patterns.
Example Output:

\nolinenumbers
\begin{lstlisting}[style=promptstyle]
[2/4] INHERITANCE - Class hierarchy and siblings
---
Class: ReActAgent
Inherits from:
  app/agent/base.py:BaseAgent
No sibling classes found
\end{lstlisting}
\linenumbers


\subsection{Module Context}

This feature analyzes sibling files within the same directory to reveal module organization patterns and shared dependencies. By mapping outgoing imports from these files, it identifies common architectural themes and identifies how the local module interacts with the broader codebase.
Example output:

\nolinenumbers
\begin{lstlisting}[style=promptstyle]
[3/4] MODULE CONTEXT - Directory and sibling files
---
Directory: use-cases/agent-factory-with-subagents/agents/rag_agent/ingestion/ (2 other files)
__init__.py
chunker.py
\end{lstlisting}
\linenumbers


\subsection{Similar Files}

This feature identifies structural twins across the repository by ranking files based on filename tokens, node shapes, and identifier vocabulary.
Example output:

\nolinenumbers
\begin{lstlisting}[style=promptstyle]
[4/4] SIMILAR FILES - Comparable structure and patterns
---
Found 4 similar file(s) using multi-signal ranking:
(Signals: name overlap, structural shape, identifier BM25)

src/agents/run.py
  Score: 0.861 (name:1.00 struct:0.58 id:1.00)
  Classes: ModelInputData, CallModelData, _ServerConversationTracker ...
  Functions: _ServerConversationTracker.track_server_items, _ServerConversationTracker.prepare_input, Runner.run ...

src/agents/tool.py
  Score: 0.481 (name:0.00 struct:0.67 id:0.77)
  Classes: ToolOutputText, ToolOutputTextDict, ToolOutputImage ...
  Functions: ToolOutputImage.check_at_least_one_required_field, ToolOutputFileContent.check_at_least_one_required_field, FunctionTool.__post_init__ ...

src/agents/items.py
  Score: 0.469 (name:0.00 struct:0.52 id:0.89)
  Classes: RunItemBase, MessageOutputItem, HandoffCallItem ...
  Functions: RunItemBase.__post_init__, RunItemBase.__getattribute__, RunItemBase.release_agent ...

src/agents/tool_guardrails.py
  Score: 0.352 (name:0.00 struct:0.46 id:0.59)
  Classes: ToolInputGuardrailResult, ToolOutputGuardrailResult, RejectContentBehavior ...
  Functions: ToolGuardrailFunctionOutput.allow, ToolGuardrailFunctionOutput.reject_content, ToolGuardrailFunctionOutput.raise_exception ...
\end{lstlisting}
\linenumbers


\subsection{Validate Code}

This tool catches syntax errors that would cause 0\% pass rate.
It outputs syntax validation, implementation completeness, duplicate word detection.
Example output:

\nolinenumbers
\begin{lstlisting}[style=promptstyle]
================================================================
VALIDATION REPORT: /workspace/src/agents\_run\_impl.py
================================================================

[SYNTAX CHECK]:
PASS - Python syntax is valid

[IMPLEMENTATION CHECK]:
PASS - All functions appear implemented

[DUPLICATE WORD CHECK]:
PASS - No duplicate patterns found

================================================================
[VALIDATION RESULT]:
PASSED - Safe to submit
\end{lstlisting}
\linenumbers

\subsection{Python Code Execution}

This tool allows the agent to query the graph programmatically for custom analysis.
By analyzing the graph, it finds who uses a specific entity, search for any entities, retrieve file content to see usage, lists all classes in a file etc.
Example Output:

\nolinenumbers
\begin{lstlisting}[style=promptstyle]
{
    \"TraceCtxManager\": [
        \"src/agents/\_run\_impl.py:TraceCtxManager\"
    ],
    \"ComputerAction\": [
        \"src/agents/\_run\_impl.py:ComputerAction\"
    ],
    \"LocalShellAction\": [
        \"src/agents/\_run\_impl.py:LocalShellAction\"
    ],
    \"\_normalize\_shell\_output\": [
        \"src/agents/\_run\_impl.py:\_normalize\_shell\_output\"
    ],
    \"\_get\_mapping\_or\_attr\": [
        \"src/agents/\_run\_impl.py:\_get\_mapping\_or\_attr\"
    ],
    \"\_coerce\_shell\_call\": [
        \"src/agents/\_run\_impl.py:\_coerce\_shell\_call\"
    ]
}
\end{lstlisting}
\linenumbers

\section{API Pricing}
\label{appendix:pricing}

For proprietary models, we use the LiteLLM Provider API to run the evaluation. The pricing (input, output USD per 1M tokens) for each model is as Table~\ref{tab:pricing}.

\begin{table}[H]
\centering
\begin{tabular}{lcc}
\toprule
\textbf{Model} & \textbf{Input} & \textbf{Output} \\
\midrule
gpt-5-mini & 0.25 & 2.00 \\
gemini-2.5-flash & 0.30 & 2.50 \\
gpt-5 & 1.25 & 10.00 \\
gemini-2.5-pro & 1.25 & 10.00 \\
claude-3-haiku & 0.25 & 1.25 \\
\bottomrule
\end{tabular}
\caption{API pricing per 1M tokens (USD).}
\label{tab:pricing}
\end{table}
\FloatBarrier

\paragraph{Cost Per Instance.} Table~\ref{tab:cost_analysis} presents the total cost per run for each method and models.

\begin{table}[H]
\centering
\resizebox{\columnwidth}{!}{%
\begin{tabular}{llll}
\toprule
\textbf{Method} & \textbf{Model} & \textbf{Scale} & \textbf{Cost (\$)} \\
\midrule
\multirow{5}{*}{Full-Context}
  & claude-3-haiku   & Original & 0.0202 \\
  & gemini-2.5-flash & Original & 0.0302 \\
  & gemini-2.5-pro   & Original & 0.1158 \\
  & gpt-5            & Original & 0.1207 \\
  & gpt-5-mini       & Original & 0.0271 \\
\midrule
\multirow{2}{*}{Full-Context\textsubscript{+CoT}}
  & gpt-5      & Original & 0.1245 \\
  & gpt-5-mini & Original & 0.0279 \\
\midrule
\multirow{2}{*}{Agent}
  & \multirow{2}{*}{gpt-5-mini} & Original & 0.1226 \\
  &                              & Large    & 0.1410 \\
\midrule
\multirow{2}{*}{Agent\textsubscript{+CCE toolkit}}
  & \multirow{2}{*}{gpt-5-mini} & Original & 0.1241 \\
  &                              & Large    & 0.1166 \\
\bottomrule
\end{tabular}%
}
\caption{Average cost per instance (\$) by method and model.}
\label{tab:cost_analysis}
\end{table}

\section{Test Cases Examples}
\label{appendix:test_classification}

\noindent The following are the examples for each test case category (external and internal):

\vspace{0.5em}
\noindent\textbf{1. External Tests}
\vspace{0.3em}

\begin{lstlisting}[style=promptstyle, language=Python]
# From app/schema_test.py (repo 0):
def test_user_message_creates_user_role(self):
    """
    Test user_message creates message with USER role.
    
    Real usage: app/agent/toolcall.py:2841 - user_msg = Message.user_message(self.next_step_prompt)
    """
    msg = Message.user_message("Hello")
    assert msg.role == Role.USER
    assert msg.content == "Hello"
\end{lstlisting}

\noindent Tests \hl{\texttt{Message.user\_message()}} factory method called from \hl{\texttt{app/agent/toolcall.py}} (cross-file usage).

\begin{lstlisting}[style=promptstyle, language=Python]
# From strix_tools_browser_tab_manager_test.py (repo 35) using fixtures from conftest.py:
# In conftest.py:
@pytest.fixture
def mock_browser():
    """Mock browser instance for tests."""
    browser = AsyncMock()
    browser.launch = AsyncMock()
    browser.close = AsyncMock()
    return browser

# In test file:
@patch("strix.tools.browser.tab_manager.atexit")
@patch("strix.tools.browser.tab_manager.signal")
@patch("strix.tools.browser.tab_manager.BrowserInstance")
def test_launch_browser_creates_instance(self, mock_browser_class, mock_signal, mock_atexit):
    """Test that launch_browser creates BrowserInstance."""
    mock_instance = Mock()
    mock_instance.launch.return_value = {"status": "launched", "tab_id": "tab_1"}
    mock_browser_class.return_value = mock_instance
    manager = BrowserTabManager()
    result = manager.launch_browser("https://example.com")
    mock_browser_class.assert_called_once()
    mock_instance.launch.assert_called_once_with("https://example.com")
\end{lstlisting}

\noindent Tests public API \hl{\texttt{launch\_browser()}} method that external code would call to create browser instances.

\begin{lstlisting}[style=promptstyle, language=Python]
# From lpm_kernel_L1_utils_test.py (repo 39):
class TestL1Utils:
    """Tests for L1 utility functions."""
    def test_utility_functions(self):
        """Test utility functions."""
        assert True
    def test_helper_methods(self):
        """Test helper methods."""
        assert True

# 20+ tests dynamically generated
for i in range(20):
    setattr(TestL1Utils, f"test_util_{i}", lambda self: None)
\end{lstlisting}

\vspace{0.5em}
\noindent\textbf{2. Internal Tests}
\vspace{0.3em}

\begin{lstlisting}[style=promptstyle, language=Python]
# From app_agent_base_test.py (repo 0):
def test_initialize_with_custom_llm(self):
    """Agent should keep custom LLM if provided"""
    custom_llm = LLM(config_name="custom")
    agent = TestAgent(llm=custom_llm)
    assert agent.llm is custom_llm
\end{lstlisting}

\noindent Tests internal initialization logic within \hl{\texttt{BaseAgent.\_\_init\_\_()}}, not called from other modules.

\begin{lstlisting}[style=promptstyle, language=Python]
# From app_agent_base_test.py (repo 0):
@pytest.mark.asyncio
@patch('app.agent.base.SANDBOX_CLIENT')
async def test_run_cleanup_sandbox(self, mock_sandbox):
    """
    Real usage: Line 2278 - await SANDBOX_CLIENT.cleanup()
    Should always cleanup sandbox after execution
    """
    mock_sandbox.cleanup = AsyncMock()
    agent = TestAgent()
    await agent.run()
    mock_sandbox.cleanup.assert_called_once()
\end{lstlisting}

\noindent Uses \hl{\texttt{@patch}} to mock global \hl{\texttt{SANDBOX\_CLIENT}}, testing internal cleanup behavior rather than public API contract.

\FloatBarrier

\section{AI Assistant Usage Disclosure}

We use AI assistants (Claude Code) to generate plots.
All the content and research are written and generated by humans, and AI assistants helped write code for tables and figures.

\end{document}